\documentclass[aps,pre,twocolumn,float,footinbib]{revtex4}

\usepackage{amsmath,amsfonts,bm,epsfig}

\usepackage{verbatim,hyperref}

\def \ed {\end{document}}
\def\Fbox#1{\vskip1ex\hbox to 8.5cm{\hfil\fboxsep0.3cm\fbox{%
  \parbox{8.0cm}{#1}}\hfil}\vskip1ex\noindent}  

\newcommand{\eq}[1]{(\ref{#1})}
\newcommand{\Eq}[1]{Eq.~(\ref{#1})}
\newcommand{\Eqs}[1]{Eqs.~(\ref{#1})}
\newcommand{\Fig}[1]{Fig.~\ref{#1}}
\newcommand{\Sec}[1]{Sec.~\ref{#1}}
\newcommand{\Ref}[1]{Ref.~\cite{#1}}
\newcommand{\Refs}[1]{Refs.~\cite{#1}}

\def\be{\begin{equation}}\def\ee{\end{equation}}
\def\bea{\begin{eqnarray}}\def\eea{\end{eqnarray}}
\def\bse{\begin{subequations}}\def\ese{\end{subequations}}
\newcommand{\BE}[1]{\begin{equation}\label{#1}}
\newcommand{\BEA}[1]{\begin{eqnarray}\label{#1}}
\newcommand{\BSE}[1]{\begin{subequations}\label{#1}}

\let \nn  \nonumber  \newcommand{\br}{\\ \nn}
\newcommand{\BR}[1]{\\ \label{#1}}

\let \= \equiv \let\*\cdot \let\~\widetilde \let\^\widehat \let\-\overline
\let\p\partial

 \def\1{\bm1} 

\def\<{\left\langle}    \def\>{\right\rangle}
\def\({\left(}          \def\){\right)}
\def \[ {\left [} \def \] {\right ]}

\renewcommand{\a}{\alpha}\renewcommand{\b}{\beta}\newcommand{\g}{\gamma}
\renewcommand{\d}{\delta}
\newcommand{\D}{\Omega}
\newcommand{\ve}{\varepsilon}
\renewcommand{\o}{\omega} 
\renewcommand{\L}{\Lambda}

\newcommand{\B}[1]{{\bm{#1}}}
\newcommand{\C}[1]{{\mathcal{#1}}}    

\renewcommand{\sb}[1]{_{\text {#1}}}  
\newcommand{\Sp}[1]{^{^{\text {#1}}}} 
\def\Sb#1{_{\scriptscriptstyle\rm{#1}}}

\begin{document}
\title{
The interaction of Kelvin waves and the non-locality of the energy transfer   in
superfluids}
\author{Jason Laurie$^\dag$}
\author{Victor S. L'vov$^*$}

\author{Sergey Nazarenko$^\dag$}

\author{Oleksii Rudenko$^*$}
\affiliation{$^\dag$~Mathematics Institute, University of Warwick, Coventry CV4 7AL, United Kingdom
}
    \affiliation{$^*$~Department of Chemical Physics, The Weizmann Institute of Science, Rehovot 76100, Israel}

\begin{abstract} We argue that the physics of interacting Kelvin Waves (KWs) is highly non-trivial and
cannot be understood on the basis of pure dimensional reasoning.
   A consistent theory of  KW turbulence in superfluids should be based upon explicit knowledge of their interactions.  To achieve this, we present
    a  detailed calculation and comprehensive analysis of the interaction coefficients for KW turbuelence, thereby, resolving previous mistakes stemming from  unaccounted contributions.
   {As a first application of this analysis, we derive a new \emph{Local Nonlinear} (partial differential)  \emph{Equation}. This equation is much simpler for analysis and numerical simulations of KWs than the Biot-Savart equation, and in contrast to the completely integrable Local Induction Approximation (in which the energy exchange between KWs is absent), describes the nonlinear dynamics of KWs.
   Secondly, we show }that the previously suggested  Kozik-Svistunov
    energy spectrum for KWs, which has often been used in the analysis of experimental and numerical data in superfluid turbulence, is irrelevant, because it is
     based upon an erroneous assumption of the  locality of the energy transfer through scales. Moreover, we demonstrate the weak non-locality of the inverse cascade spectrum with a constant  particle-number flux and find resulting logarithmic corrections to this spectrum.
\end{abstract}


\maketitle


\section*{\label{s:intro}Physical background, {methodology}  and overview of results}
\subsection{\label{ss:KW}Kelvin waves (KWs) in superfluid turbulence}
The role of Kelvin Waves (KWs) in the dissipation of energy in zero temperature quantum turbulence   has long been discussed within the quantum turbulence community.  It is widely believed that KWs extend the transfer of a constant energy flux from the fully $3$D Kolmogorov-like turbulence at large scales, through a crossover mechanism at scales comparable to the inter-vortex distance, to smaller scales via a local KW cascade on quantized vortices.  Much theoretical work has been done recently, including the conjecture of a power-law scaling for the KW cascade made by Kozik and Svistunov  in 2004, the KS-spectrum~\cite{KS04}.

Nevertheless there remain important unanswered questions in quantum turbulence: \\
-- What are the relative roles of KWs and the other processes, e.g. vortex reconnections,
in the transfer of energy to small scales?\\
-- What are the dominant physical mechanisms in the classical-quantum crossover range? Two alternative scenarios were put
forward for this range: firstly, one relying on the idea that the polarization of vortex tangles suppress vortex reconnections, which
lead to a bottleneck hump \cite{LNR-1,LNR-2}, and the second, implies that reconnections play an active role in removing the bottleneck
\cite{kozlik08}. \\
--  If the KWs do play a key role at small scales, what kind of interaction processes are important for the transfer of energy towards smaller scales?
Is it the resonant wave-wave interactions, or a linear process of wave number evolution due to a large-scale curvature and/or
slow time dependence of the underlying vortex line, or any other possibility? \\
-- If the KW energy transfer is dominated by the six-wave scattering, can one safely assume, as in~\cite{KS04},  that this process  is local,  in the sense that the $\B k$-waves (with a given wave vector $\B k$) are
  mainly affected by  $\B k'$-waves {(with a given wave vector $\B k'$, where $k'=|\B k'|$ is of} the same order as $k=|\B k|$, with contributions of $\B  k'$-waves with $k'\ll k$ and $k'\gg k$  being vanishingly small?

In this paper we   {do not}
address all of these problems, particularly the {ones} about the role of reconnections and about
the structure of the crossover range. We restrict our attention to the  nonlinear
interactions of weakly nonlinear KWs propagating on a single straight vortex line.
This corresponds to the small-scale range of superfluid turbulence at near-zero temperature, where, because of {the} short
wavelengths of KWs, one can ignore {the} influence of the neighboring vortex lines {with}in the tangle, and assume that nonlinearity,
being weak, is {still} strong enough for the nonlinear evolution to proceed faster than the large-scale (space and time) changes in the
underlying vortex line.

Within this idealized setup, our immediate goal {is to} revise and advance
 the theory of weak-wave turbulence of KWs. 
In particular, we have clarified the structure of the nonlinear KW interactions, corrected the theory by including previosuly unaccounted leading-order contributions to the effective wave Hamiltonian, and explicitly calculated the interaction coefficients.  Furthermore, we have used these results to:\\
-- Firstly, derive a simple local nonlinear equation~\eq{LNE} for describing KW turbulence.\\
-- Secondly, to check locality of the KS-spectrum assumed in the previous theory.

\subsection{\label{ss:mo} Hierarchy of the  equations of motion for KWs}

A commonly accepted  model of superfluid turbulence  comprises a randomly moving tangle of quantized vortex lines which can be characterized by the mean intervortex distance $\ell$ and the vortex core radius $a_0 \ll \ell$.
   There are two approaches {in dealing} with the vortex core. {The} first one is a ``microscopic" model, in which the core is resolved:
   it is based on the Gross-Pitaevski equation,
   \be
   \frac{\partial \Psi }{\partial t} + \nabla^2 \Psi - \Psi |\Psi|^2 =0,
   \ee
   where $\Psi$ is the so-called condensate wave function.


This model was systematically derived for the Bose-Einstein condensate
   in super-cold atoms, and not for liquid Helium.
  Nevertheless, it is frequently used for describing superfluid flows in Helium
because it contains several essential features of such superfluids, i.e. vortex quantization, acoustic waves (phonons) in {the} presence
 of a condensate, and the {description of} a gradual (nonsingular) vortex line reconnection.

However, the    Gross-Pitaevski equation can be costly to study, and one often resorts to using the so-called Biot-Savart
formulation of the Euler equations for ideal classical fluids, exploiting the fact
that far away from the vortex cores the Gross-Pitaevski dynamics is isomorphic to the ideal classical flow via
the Madelung transformation. In the Biot-Savart model, the vortices are postulated by a cutoff in the
equations for the vortex line elements. Namely, the equations used are \cite{1,2,AH65}:
\be
\label{biot-savart}
\dot{\bf r} = \frac{\kappa}{4 \pi} \int {d {\bf s} \times ({\bf r}-{\bf s})
\over |{\bf r}-{\bf s}|^3},
\ee
with a cutoff
at the core radius $a_0$, i.e. integration is over the range $|{\bf r}-{\bf s}|>a_0$. Here $\kappa \= {2\pi \hbar}/m$ is the quantum of
velocity circulation,  $m$ is the particle mass. In what follows, we will adopt the Biot-Savart equation ({BSE})  
as a staring point for our {derivation}. 

To  consider the KW system, one has to start with
an equilibrium state corresponding to an infinitely long straight vortex line and
perturb it with small angle disturbances. This corresponds to a setup of weakly nonlinear
KWs which are dispersive, and that can be decribed by weak-wave turbulence theory.
 For this, one has to parametrize the transverse displacement vector $(x,y)$ of the perturbed line
 by the  distance $z$ along the unperturbed line (the latter lies along the Cartesian $z$-axis):
 \be
 w(z,t)\= x(z,t)+
i\,y(z,t).
 \ee
 In terms of $w$, the
BSE can be written in a Hamiltonian form
\BE{Ha}
i\kappa\,\frac {\p w}{\p t} = \frac{\delta H\{w,w^*\} }{ \delta w^*}\,,
\ee
where $\d / \d {w}^*$ is the functional derivative, the asterisk stands for the complex conjugation  and the Hamiltonian $H= H\Sp{\,BSE}$ is defined as follows \cite{S95},
\begin{equation}
H\Sp{\,BSE}\! = {\kappa \over 4 \pi} \int\!\!\!\!\int\! {\,\big[1 + Re\big(w'^{*}(z_1) w'(z_2)\big)\big]\,dz_1 dz_2\,
\over \sqrt{(z_1 - z_2)^2 + |w(z_1)-w(z_2)|^2\,}},
\label{BSE}
\end{equation}
where we have used the notation $w'(z)=d w/ d z$.

 Then, one must   expand in two small parameters:
the  perturbation inclination $w' \ll 1$ and $\Lambda_0^{-1} \ll 1$, where
$\Lambda_0=\ln (\ell/a_0)$.
Such a simultaneous expansion is not easy. This is because the leading order
  in $1/\Lambda_0$ gives an integrable model called the Local Induction Approximation (LIA) \cite{AH65,S95}
  with  a  Hamiltonian
  \begin{equation}
H\Sp{LIA} =  {\kappa \over 2 \pi} \L_0
\int dz \sqrt{1 + |w'(z)|^2}.
\label{LIA}
\end{equation}
Because of the integrability, LIA conserves an infinite number of integrals of motion, and wave resonances
are {absent} in all orders, which prevents energy {exchange between KWs}.
Thus, LIA appears to be too simple, and to describe the KW energy transfer, one has to go to next order in $1/\Lambda_0$.

    The second difficulty is that the lowest order process, the four-wave resonances, are absent for
    such one dimensional systems with concaved dispersion relations. Thus, one must also go to the
    next order in small $w'$.
The combination of these two facts makes finding the effective interaction Hamiltonian
${\C H}_{\hbox{\small int}}$  for KWs a hard task.

For numerical analysis {purposes}, the BSE model~\eq{biot-savart} is also quite challenging because of the nonlocal (in the physical
space) integral that has to be computed, especially when one has to resolve a wide range of turbulent scales
and when the waves are weak, so that the evolution times are long.


Thus, there is a clear need for a simpler model for nonlinear KWs, which would be local in the
physical space like LIA (i.e. represent a nonlinear partial differential equation), but unlike LIA would be
capable of descripting the energy transfer over turbulent scales.
 Motivated by this need, an \em ad hoc \em model was introduced in
 \cite{boffetta} which has the simplest possible form with all the scaling properties and solutions
 of the original BSE model preserved. This model was called the Truncated-LIA (or TLIA); it has the following Hamiltonian
 \begin{equation}
\label{tLIA}
 H\Sp{TLIA}= \frac{\kappa^2}{4\pi}\,\L_0\!
 \int \left [   |w'|^2 -
 \frac{1}{4}  |w'|^4 \right ]d z   \,,
 \end{equation}
 The name TLIA arises from the fact that it can formally be obtained by
 expanding the LIA Hamiltonian in $|w'|^2$ and truncating at the fourth order.
  This truncation leads to the breaking of the LIA integrability, while preserving all
 the important scalings.

 The   TLIA model turned out to be very efficient and useful for numerical simulations \cite{boffetta},
 even though it was suggested \em ad hoc \em , and motivated by the need for greater simplicity, rather
 than derived from first principles.
 In the present paper, we will obtain a very simple, local in $z$, nonlinear partial differential equation which is isomorphous to
 TLIA for weakly nonlinear KWs, and which is obtained rigorously from BSE by asymptotic expansions
 in  $1/\Lambda_0$ and $w'$, and by a subsequent identification of the dominant type of interaction wave sextets.

\subsection{\label{ss:new} The new Local-Nonlinear Equation for Kelvin waves}

Our main goal in the paper is, being based on the BSE,  is to systematically derive
an effective motion equation for KWs, ``as simple as possible, but not more". {From one side, this new equation will include dynamical interaction (with energy exchange) of KWs (which are absent in LIA), and from the other, will contain the leading contribution obtained from local elements (in contrast to the TLIA model).  Moreover, at the same time, being drastically simplier than the BSE itself.}   The resulting equation reads:
\Fbox{\bea\label{LNE}  i \, \frac{\p  \widetilde{w}}{\p t}+ \frac{\kappa}{4\pi}\, \frac{\p}{\p z}\left[ \left(\L - \frac 14 \, \left | \frac{\p \widetilde{w} }{\p z }\right |^4 \right) \frac{\p  \widetilde{w} }{\p z } \right]=0\,. ~~~~ \\ \nn
\mbox{Local Nonlinear Equation (LNE) for KWs.}~~~  \eea}
  The variable $\widetilde{w}(z,t)$ is related to
$w(z,t)$ via a weakly nonlinear canonical transformation of type
\be\label{ct}
\~w(z,t)=w(z,t)+ O({w^\prime}^{\,3})\ .
\ee
The dimensionless parameter
\be \label{WWb}
 \L \= \L _0-\gamma - \frac32\simeq \L _0-2\,,
\ee
where $\gamma=0.5772\dots$ is  the Euler constant. 
The replacement $\L_0\to \L$  is equivalent to a replacement of $a_0$  by an effective vortex core radius  $a=a_0\,  \exp(\gamma+3/2) \simeq 8 \, a_0$, in the equation $\L=\ln (\ell/a)$.

We entitled the result (\ref{LNE}) as the \emph{Local Nonlinear Equation} (LNE) to stress its three main features:
\begin{description}
\item \emph{ Locality}  (in physical space) of the KW interactions:  according to \Eq{LNE}, the evolution of the KW amplitude  $\~w(z, t)$ depends only on the  slope $\p \~ w (z,t)/ \p z$ of KWs  and on its curvature (via $\p^2 \~ w (z,t)/ \p z^2$) at the same point $z$.

\item   \emph{Nonlinearity}  plays a crucial role in the LNE~\eq{LNE}, as it is responsible for  the energy transfer among the KW modes. Notice that LIA, being formally a nonlinear equation,  allows (in the framework of the inverse scattering formulation) a linear formulation, in which the energy exchange between KWs is explicitly absent.

\item  The word \emph{Equation} (not an  ``Approximation") stresses  the fact that LNE~\eq{LNE}   is asymptotically exact   in the triple limit
\be \label{WWc}
w^{\,\prime} \ll 1\,, \quad \L^{-1}\ll 1\,, \quad \mbox{and} \quad \lambda \ll \ell\,,
\ee
where $\lambda$ is the characteristic wavelength of the the KWs.
\end{description}

It is instructive to represent LNE~\eq{LNE}  in a Hamiltonian form
\bse \be\label{WWd}
i\, \kappa\, \frac{\p \widetilde{w}}{\p t}= \frac {\d}{\d \widetilde{w}^*}\,H\Sp{\,LNE}\{\widetilde{w}, \widetilde{w}^*\}\,,
\ee
where 
  \Fbox {\bea\label{WWe}
 H\Sp{\,LNE}= \frac {\kappa^2}{4\pi}    \int{\! \left[\,{\Lambda} \left |\frac{\p \widetilde{w}}{\p z}\right |^2\! - \frac{1}{12} \left |\frac{\p \widetilde{w}}{\p z} \right |^6\, \right ]d z}   \,, ~~~\\ \nn
 \mbox {LNE Hamiltonian. }~~~~~~~~~~~~~~~~~~~~~~~~~
 \eea  }
\ese
In the weakly nonlinear case, the LNE model is isomorphic to the TLIA model \eqref{tLIA},
{in which}
the variable $w$ is related to $\widetilde{w}$ via the canonical transformation mentioned above
and a proper rescaling.
We repeat that the TLIA model was introduced in \cite{boffetta} in an  \em ad-hoc \em
way, requiring that it must have the same basic scaling properties as the original BSE \eqref{biot-savart}, but be
simpler than the latter. In the present paper, the LNE~\eq{LNE} is derived from the BSE \em systematically \em
and, therefore, it is an asymptotically rigorous equation in the relevant KW turbulence limits.
In spite of the remarkable simplicity  of the LNE~\eq{LNE}, its asymptotical derivation is rather cumbersome, which possibly explains  the fact that it has never been obtained before. {To reach our goals} we proceeded as follows.

\subsection{\label{ss:nonloc} Challenges of the KW turbulence. Nonlocality of the six-wave  process.}

  The first part of the paper is devoted to an explicit calculation of the effective six-wave interaction coefficient in the
limit of small perturbation angles and small values of $1/\Lambda$ {that culminate in the derivation of the LNE~\eq{LNE}. As we mentioned above, the} respective expansion in two small parameters is not easy and rather cumbersome.
To keep our presentation reasonably short and transparent, in \Sec{ss:Ham} we describe  only  the main steps  of the calculation and move details to the Appendices. In this part, we fix a set of rather important technical errors made in \Refs{KS04,KS09}, thereby preparing the mathematical problem for further analysis.

As {an important} 
 application of the obtained result, we dedicate \Sec{s:kin}
   to an analysis of the cascades of energy $E$ and particle-number $N$  caused by the $3  \leftrightarrow 3$-scattering of KWs in the context of the KS-conjecture~\cite{KS04}.  The KS theory predicts a power-law energy spectrum with constant $E$-flux, $E\Sb{KS}(k)\propto k^{-7/5}$, and with constant $N$-flux, $E\Sb{N}(k)\propto k^{-1}$~\cite{inverseKW}.   These spectra  can only be  valid if they are \emph{local}.
  We show  in \Sec{ss:nonloc}   that the KS spectrum is strongly non-local, so that the KW dynamics are dominated by interactions of $\B k'$-waves with $k'\ll k$. Therefore, the KS spectrum is {physically} irrelevant, i.e. it cannot be realized in Nature. In \Sec{ss:nonloc}, we also demonstrate  that the $N$-flux spectrum is weakly nonlocal in the sense that the dynamics of the $\B k$-waves are equally effected by all the $\B k'$-waves with $k'   \lesssim k$.
  In \Sec{sss:log}, we ``fix" locality of this spectrum by a logarithmic correction, $E\Sb{N}(k)\propto k^{-3}[ \ln  (k \ell)]^{-1/5}$, \Eq{corr}.

Establishing nonlocality of the six-wave theory is a key step to the desired effective description. Indeed, it suggests that
the dominant wave sextets must involve modes with widely separated wavelengths.
We will show that in this case, the effective interaction coefficient looks remarkably simple: it is proportional to the product of the six
wave-numbers of the sextet modes, see \eqref{ASa}. This immediately yields the LNE~\eq{LNE}.
In conclusion, we discuss prospects of building a theory based upon the LNE~\eq{LNE}, which should lead to an alternative result to the
invalid KS spectrum.

\section{\label{s:stat}Hamiltonian Dynamics  of KWs}

\subsection{Introduction to the problem}
To derive an effective KW Hamiltonian leading to the LNE~\eq{LNE}, we first briefly overview the Hamiltonian description of KWs initiated in~\cite{KS04} and further developed in \cite{LNR-1,boffetta}. The main goal of \Sec{ss:Ham} is to start with the so-called ``bare" Hamiltonian \eq{BSE} for the Biot-Savart description of KWs~\eq{biot-savart}  and obtain expressions for the  frequency~\Eqs{Ok},    four- and six-KWs interaction coefficients~\Eqs{T-intro} and \eq{W-intro} and their $1/\L$-expansions, which will be used in further analysis.
\Eqs{Ok}, \eq{T-intro} and \eq{W-intro} are  starting points for further modification of the KW description, given in \Sec{ss:trans}, in which we explore the consequences of the fact that \emph{non-trivial} four-wave interactions of KWs are prohibited by the conservation laws of energy and momentum:
\be\label{4-KW}
\o_1+ \o_2=\o_3+ \o_4\,, \qquad  \B k_1+\B k_2=\B k_3+\B k_4\,,
\ee
where $\o_j\=\o(k_j)$ is the frequency of the $k_j$-wave. Only the \emph{trivial} processes with $\B k_1=\B k_3$,   $\B k_2=\B k_4$, or $\B k_1=\B k_4$,   $\B k_2=\B k_3$ are allowed.

\begin{figure}
 \includegraphics[width=3.5cm]{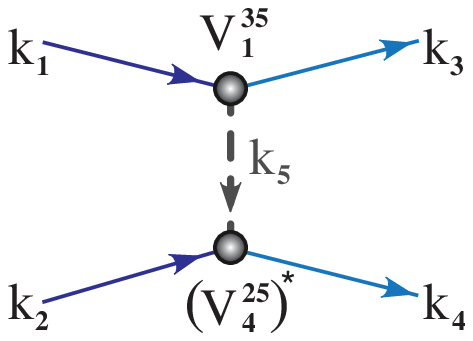} ~~~~~~ \includegraphics[width=3.5cm]{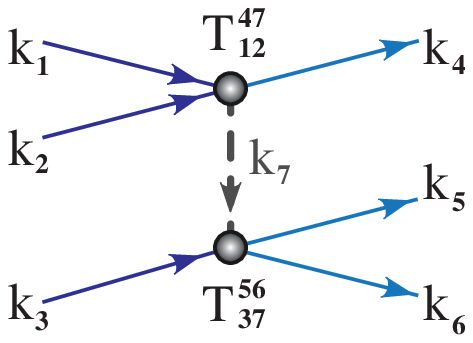}
\caption{\label{f:1}{ Examples of  contribution of the triple vertices to the four-wave Hamiltonian when three-wave resonances are forbidden (left) and  contributions of the quartet vertices to the six-wave Hamiltonian in resonances when four-wave resonances are forbidden (right).
 Intermediate virtual waves are shown by dash lines.} }
\end{figure}

It is well known (see, e.g. \Ref{ZLF}) that in the case when nonlinear wave processes of the same kind (e.g. $1 \rightarrow 2$) are \emph{forbidden} by conservation laws, the terms corresponding to this kind of processes can be eliminated from the interaction Hamiltonian by a  weakly nonlinear canonical transformation.
A famous example \cite{ZLF} of this procedure comes from a system of gravity waves on water surface in which
    three-wave resonances
$\o_1 = \o_2+\o_3$
 are forbidden. Then, by a canonical transformation to a new variable $b = a + O (a^2)$, the old Hamiltonian $H(a, a^*)$ is transformed to a new one, $\~H(b, b^*)$ , where the three-wave (cubic) interaction coefficient $V_{\ 1}^{2,3}$ is eliminated at the expanse of appearance of an additional contribution to the next order term (i.e. four-wave interaction coefficient $T_{\, 1,2}^{\,3,4}$) of the type 
\be 
    \label{34} 
    V_{\ 1}^{3,5} \big(V_{\ 4}^{2,5}\big)^* \big/ [\,\o_5+\o_3-\o_1\,]\ .
\ee 
One can consider this contribution as a result of the second-order perturbation approach in the three-wave processes $ \B k_1\rightarrow  \B k_3 + \B k_5$ and $ \B k_5 +  \B k_2 \rightarrow  \B k_4$, see \Fig{f:1}, left. The virtual wave $ \B k_5$  oscillates with a forced frequency   $\o_1-\o_3$ which is \emph{different} from its eigenfrequency $\o_5$. The  inequality  $\o(k_1)-\o(k_3)\ne \o(| \B k_1- \B k_3 |)$ is a consequence of the fact that the three-wave processes  $\o(k_1)-\o(k_3)=\o(| \B k_1- \B k_3 |)$ are forbidden. As the result the denominator in \Eq{34} is non-zero and the perturbation approach leading to \Eq{34} is applicable when the waves' amplitudes are small.

Strictly speaking our problem is different:  as we mentioned above, not all $2 \leftrightarrow 2$ processes~\eq{4-KW} are forbidden, but only the non-trivial ones that lead to energy exchange between KWs. Therefore, the use of a weakly nonlinear canonical transformation~\eq{ct} (as suggested in \cite{KS04}) should be done with extra caution. The transformation~\eq{ct} is supposed to eliminate the fourth order terms from the BSE-based interaction Hamiltonian by the price of appearance of extra contributions to the ``full" six-wave interaction amplitude $\widetilde{W}_{\ 1,2,3}^{\,4,5,6}$, \eqref{tW-exp},  of the following  type (see \Fig{f:1}, {right}):
\be
    \label{46}
   \frac{ T_{\, 1,2}^{\,4,7} \ T_{\ 3,7}^{\,5,6}}{\,\o_7+\o_4-\o_1-\o_2}\,, \quad \o_7\=\o(|\B k_1+\B k_2-\B k_4|)\ .
\ee
Here  all wave vectors are taken on the six-wave resonant manifold
\bea\label{6-KW}
\o_1+ \o_2+\o_3&=& \o_4+\o_5+\o_6\,, \br
 \B k_1+\B k_2+\B k_3&=&\B k_4+ \B k_5 + \B k_6\ .
\eea
The danger is seen from a particular example when $\B k_1\to \B k_4\,, \ \B k_2\to \B k_5$ and $\bm {k}_3\to \B k_6$, so $\o_7\to \o_2$, and the denominator of \Eq{46} goes to zero, while the numerator remains finite. This means, that the perturbation contribution~\eq{46} diverges and this approach becomes questionable.

However a detailed analysis of \emph{all} contributions of the type \eq{46} performed a-posteriori and presented in \Sec{ss:trans}  demonstrates cancelations of diverging terms with opposite signs such that the resulting ``full" six-wave interaction coefficient remains finite, and the perturbation approach~\eq{46} {appears} to be eligible. The reason for this cancelation is {hidden deep within the} symmetry of the problem, which will not be discussed here.

Moreover, finding the ``full" Hamiltonian is not enough for formulating the effective model.
Such a model must include all contributions in the same $\mathcal{O}(1/\L)$, namely the
leading order allowing {energy} transfer in the $\B k$-space.
The ``full" Hamiltonian still contains un-expanded in ($1/\L$) expressions for the
KW frequencies. Leaving only the leading (LIA) contribution in the KW frequency $\o$,
as it was done in \cite{KS04,KS09}, leads to a serious omission of an important leading order
contribution. Indeed, the first sub-leading contribution to $\o$ shifts the LIA resonant manifold,
which upsets the integrability.
As a result, the LIA part of  $\widetilde{W}_{\ 1,2,3}^{\,4,5,6}$ yields a contribution to the  effective model
in the leading
order. This (previously overlooked) contribution will be found and analyzed in
\Sec{ss:effD}.

\subsection{\label{ss:Ham} ``Bare" Hamiltonian dynamics of KWs}
\subsubsection{\label{sss:can}Canonical form of the ``bare" KW Hamiltonian}

Let us postulate that the motion
of a tangle of quantized vortex lines can be described by
the BSE model \eqref{biot-savart}, and assume that
\begin{eqnarray}
      \label{Lambda-def}
      && \Lambda_0 \= \ln\!\big(\ell/a_0\big) \gg 1, 
\end{eqnarray}
where $a_0$ is the vortex core radius.
The BSE can be written in the Hamiltonian form \eqref{Ha} with Hamiltonian \eqref{BSE}.
 Without the
cut-off, the integral in $H\Sp{BSE}\!$, \Eq{BSE}, would be
logarithmically divergent with the dominant contribution given by
the leading order expansion of the integrand in small $z_1-z_2$,
that corresponds to $H\Sp{LIA}\!$, \Eq{LIA}.

As we have already mentioned, LIA represents a completely integrable system
and it can be reduced to the one-dimensional nonlinear Schr\"{o}dinger (NLS)
equation by the Hasimoto transformation \cite{hasimoto}.
However, it is the complete integrability of LIA that makes it insufficient for
describing the energy cascade and which makes it necessary to
consider the next order corrections within the BSE model.

For small-amplitude KWs when  $ w'(z)\ll 1 $, we can expand the
Hamiltonian \eqref{BSE} in powers of ${w'}^{\,2}$, see Appendix~\ref{aa:calc}:
\BE{exp} H=H_2+H_4+H_6+ \ldots
\ . \ee
  Here we omitted the constant term $H_0$ which does not contribute to the equation of motion \Eq{Ha}.
  Assuming that the  boundary conditions are periodical on the length $\C L$ (the limit $k\C L\gg 1$ to be taken later) we can   use  the
Fourier representation
\BE{Four}
    w(z,t)= \kappa^{-1/2} \sum_k  a(\B k,t) \exp (i\B kz)\,. 
\ee
The bold face notation of the one-dimensional wave vector is used for convenience only. Indeed, such a vector is just a real number, $\B k \in \mathbb{R}$, in our case for KWs. For further convenience, we reserve the normal face notation for the length of the one-dimensional wave vector, i.e. $k=|\B k | \in \mathbb{R}_{\B{+}}$.
In Fourier space,  the Hamiltonian equation also  takes  a
canonical form:
\BSE{Ham} \BE{can}
i\,   \frac{\p a(\B k,t)}{\p t } = \frac{\delta  \C H\{a,a^*\} }{\delta  a^*(\B k,t)}\ .
\ee
  The new Hamiltonian $\C H$ is the density of the old one:
\BE{HamA} \C  H\{a ,a ^*\}=  H\{w,w^* \} / \C L = \C H_2 + \C H_4 + \C H_6 + \dots
\ee
The Hamiltonian
\BE{Ham2} 
  \C H_2 = \sum_k  \o_k\, a_k\, a^*_k \,,
\ee
describes the free propagation of linear KWs with the dispersion law   $\o_k\=\o(k)$, given by \Eq{Ok},  and
the canonical amplitude $a_k\=a(\B  k,t)$.
The interaction Hamiltonians $\C H_4$ and  $\C H_6$ describe the four-wave processes of $2 \leftrightarrow 2$ scattering and the six-wave processes of $3 \leftrightarrow 3$ scattering respectively.
Using a short-hand notation $a_j\=a(\B k_j,t)$, they can be written as follows:  
\BEA{Ham4}
      \C H_4 &=& \frac14 \sum_{1+2=3+4} T_{\,1,2}^{3,4}\ a_1a_2a_3^* a_4^* \,, 
  \\
    \C H_6 &=& \frac1{36} \sum_{1+2+3=4+5+6}\!\!\!  W_{\,1,2,3}^{4,5,6}\ a_1a_2a_3 a_4^*a_5^* a_6^*\ .
\eea
\ese
Here   $T_{\,1,2}^{3,4} \= T(\B k_1,\B k_2|\B k_3,\B k_4)$  and  $W_{\,1,2,3}^{4,5,6} \= W(\B k_1,\B k_2,\B k_3|\B k_4,\B k_5,\B k_6)$ are ``bare" four- and six-wave interaction coefficients, respectively. Summations over $\B k_1\dots \B k_4$ in $\C H_4$ and over $\B k_1\dots\B  k_6$ in $\C H_6$ are constrained by  $\B k_1+ \B  k_2 = \B  k_3+\B k_4$ and by $\B k_1+ \B k_2+ \B  k_3 =\B  k_4+\B k_5+\B k_6$, respectively.

\subsubsection{\label{sss:fun}$\L$-expansion of the bare Hamiltonian function}
As will be seen below, the leading terms in the Hamiltonian functions $\o_k$, $T_{\,1,2}^{3,4}$  and $W_{\,1,2,3}^{4,5,6}$, are proportional to $\L$, which correspond to the LIA \eqref{LIA}.  They will be denoted  further  by a front superscript ``~$^\Lambda$~", e.g. $^\L \o_k$, etc.  Because of the  complete integrability, there are no dynamics in the LIA.  Therefore, the most important terms for us will be the ones of zeroth order in $\Lambda$, i.e.
the ones proportional to $ \L^0 = \C O(1)$. These  will be denoted by a front superscript ``~$^{1}$~" , e.g. $^1 \o_k$, etc.

Explicit calculations of the Hamiltonian coefficients must be done very carefully, because even a minor mistake in the numerical prefactor can destroy various cancelations of  large terms in the Hamiltonian coefficients. This could change the order of magnitude of the answers and the character of their dependence on the wave vectors in the asymptotical regimes. Details of these calculations are presented in Appendix~\ref{aa:calc}, whereas the results are given below.

Together with Eq.~(\ref{WWb}), the Kelvin wave frequency is:  
\BSE{Ok}
  \BEA{OkA}
  \omega_k &=& {^\Lambda\omega}_k +{^1\omega}_k +\mathcal{O}(\Lambda^{-1})\,, \quad \mbox{where}\BR{OkB}
        {^\Lambda\omega}_k & = &\kappa\, \Lambda \,{k^2}\big/{4\pi}\,,\BR{OkC}
         {^1\omega}_k &=& -\kappa\,{k^2}  \ln\!\big(k\ell\big) \big/{4\pi}\ .
  \end{eqnarray}
\ese  

The ``bare" 4-wave interaction coefficient is:
\BSE{T-intro}
  \BEA{T-intoA}
  T_{\,1,2}^{3,4} &=& {^\Lambda T_{\,1,2}^{3,4}} +{^1 T_{\,1,2}^{3,4}} +\mathcal{O}(\Lambda^{-1}) \,, \BR{T-introB}
      {^\Lambda T_{\,1,2}^{3,4}} &=& -\Lambda\,\B k_1\B k_2\B k_3\B k_4\big/{4\pi}\,,\BR{T-introC}
    {^1 T_{\,1,2}^{3,4}} &=& -\left(5\, \B k_1\B k_2\B k_3\B k_4 +\mathcal{F}_{\,1,2}^{\,3,4}\, \right)\!\!\big/{16\pi}\ .
  \end{eqnarray}
\ese
The function $\mathcal{F}_{\,1,2}^{\,3,4}$ is symmetric with respect to $\B k_1 \leftrightarrow \B k_2$, $\B k_3 \leftrightarrow \B k_4$ and $\{\B k_1,\B k_2\} \leftrightarrow \{\B k_3,\B k_4\}$; its definition is given in Appendix~\ref{aa:F4}.

The ``bare" 6-wave interaction coefficient is \
\BSE{W-intro}
  \BEA{W-introA}
  W_{\,1,2,3}^{4,5,6} &=& {^\Lambda W_{\,1,2,3}^{4,5,6}} + {^1 W_{\,1,2,3}^{4,5,6}} +\mathcal{O}(\Lambda^{-1})  \,,  \BR{W-introB}
       {^\Lambda W_{\,1,2,3}^{4,5,6}} &=& \frac {9\, \Lambda}{8\pi\kappa}\,\,\B k_1\B k_2\B k_3\B k_4\B k_5\B k_6 \,,   \BR{W-introC}
    {^1 W_{\,1,2,3}^{4,5,6}} &=& \frac{9 }{32\pi\kappa}\big(7\, \B k_1\B k_2\B k_3\B k_4\B k_5\B k_6 -\mathcal{G}_{\,1,2,3}^{\,4,5,6}\, \big)  \ .
  \end{eqnarray}
\ese
The function $\mathcal{G}_{\,1,2,3}^{\,4,5,6}$ is symmetric with respect to $\B k_1 \leftrightarrow \B k_2 \leftrightarrow \B k_3$, $\B k_4 \leftrightarrow \B k_5 \leftrightarrow \B k_6$ and $\{\B k_1,\B k_2,\B k_3\} \leftrightarrow \{\B k_4,\B k_5,\B k_6\}$; its definition is given in Appendix \ref{aa:G6}.

Note that the full expressions for $\omega_k$, $T_{\,1,2}^{3,4}$ and $W_{\,1,2,3}^{4,5,6}$ do not contain $\ell$ but rather  $\ln{\!(1/a_0)}$.
This is natural because  in the respective expansions $\ell$ was introduced as an auxillary parameter facilitating the calculations, and it
does not necessarily have to coincide with the inter-vortex distance. More precisely, we should have used some effective
intermediate length-scale, $\ell \sb{eff}$, such that $\ell \ll  \ell \sb{eff} \ll 2\pi /k$. However, since  $ \ell \sb{eff}$  is artificial and would
have to drop from the full expressions anyway, we chose to simply write $\ell$ omitting subscript ``eff". Cancelation of $\ell$ is
a useful check for verifying the derivations.

\subsection{\label{ss:trans}Full  ``six-KW" Hamiltonian dynamics}
\subsubsection{\label{sss:fullH}Full six-wave interaction Hamiltonian $\~{\C H}_6$}
Importantly,  the four-wave dynamics in one-dimensional media  with concaved dispersion laws $\o(k)$
are absent because the conservation laws of energy and momentum allow only trivial processes with $\B k_1=\B k_3$,   $\B k_2=\B k_4$, or $\B k_1=\B k_4$,   $\B k_2=\B k_3$.
This means that by a proper nonlinear canonical transformation $\{a, a^*\} \Rightarrow \{b,b^*\}$, $\C H_4$ can be eliminated from the Hamiltonian description. This comes at a  price of  appearance of additional terms in the full interaction Hamiltonian $\widetilde{\C H}_6$:
\BSE{tHam}
\BEA{tHamA}
{\C H}\{a,a^*\}&\Rightarrow&\~{\C H}\{b,b^*\}= \~{\C H}_2+ \~{\C H}_4+ \~{\C H}_6+\dots\,, 
\BR{tHamB}
\~{\C H}_2 &=&\sum_{k}\,  \o_k\, b_k\, b^*_k \,,\qquad \~{\C H}_4\= 0\,, 
\BR{tHamC}
\~{\C H}_6 &=& {\frac1{36}} \sum_{1+2+3=4+5+6}\!\!\!  \~{W}_{\,1,2,3}^{4,5,6}\ b_1b_2b_3 b_4^*b_5^* b_6^*\,, 
\BR{tHamD}
  {\~W}_{\,1,2,3}^{4,5,6}   &=&  W_{\,1,2,3}^{4,5,6}+ Q_{\,1,2,3}^{4,5,6}\,, \BR{tHamE}
   Q_{\,1,2,3}^{4,5,6}  &=& \frac{1}{8}\!\! \sum^{3}_{\substack{i,j,m=\,1\\ i\neq j\neq m}} \sum^{6}_{\substack{p,q,r=\,4\\ p\neq q \neq r}}\!\!\! q_{\,\,i,j,m}^{\,p,q,r}\,, \br  q_{\,\,i,j,m}^{\,p,q,r}&\=&  \frac{T_{\,r,\,j+m-r}^{j,\,m}\,T_{\,i,\,p+q-i}^{q,\,p}}{\Omega_{\,j,\,m}^{\,r,\,j+m-r}}  +\frac{T_{\,m,\,q+r-m}^{q,\,r} \, T_{\,p,\,i+j-p}^{i,\,j}}{\Omega_{\,q,\,r}^{\,m,\,q+r-m}} \,, \\ 
    \label{tHamF}
  \Omega_{1,2}^{3,4} &\=& \omega_1+\omega_2 -\omega_3-\omega_4 \\
  &=& {^\Lambda\Omega}_{1,2}^{3,4} +{^1\Omega}_{1,2}^{3,4} +\mathcal{O}(\Lambda^{-1})\nonumber \ .
\eea
\ese
The $Q$-terms in the\emph{ full six-wave interaction coefficient} $ \~{W}_{\,1,2,3}^{4,5,6} $ can be understood as contributions of two four-wave scatterings into resulting six-wave process via a virtual KW with $\B k=\B k_j+\B k_m-\B k_r$ in the first term in $Q$ and via a KW with $\B k=\B k_q+\B k_r-\B k_m$ in the second term; see \Fig{f:1}, {right}.

\subsubsection{\label{sss:fun}$1/\L$-expansion of the full   interaction coefficient $ \~{W}_{\,1,2,3}^{4,5,6}$ }
Similarly to \Eq{W-introA}, we can present $ \~{W}_{\,1,2,3}^{4,5,6}$ in the $1/\L$-expanded form:
\BE{tW-exp} 
  \widetilde {W}_{\,1,2,3}^{4,5,6} =  {^\Lambda \~{W}_{\,1,2,3}^{4,5,6}}  + {^1 \~{W}_{\,1,2,3}^{4,5,6}} +\mathcal{O}(\Lambda^{-1})  \,,
\ee
Due to the complete integrability of the KW system in the LIA, even the six-wave dynamics must be absent in the interaction coefficient $^\L\~ {W}_{\,1,2,3}^{4,5,6}$.  This is true if function $^\L\~ {W}_{\,1,2,3}^{4,5,6}$  vanishes on the LIA resonant manifold: 
\BSE{LIA1}
\BEA{LIA1a}
  ^\Lambda {\widetilde{W}}_{\ 1,2,3}^{\,4,5,6}\ \d_{\,1,2,3}^{4,5,6}\ \d \big(\, {^\L \~\D _{\,1,2,3}^{\,4,5,6}}\, \big)\=0\,,
\end{eqnarray}
where
\BEA{LIA1bc}
\d_{\,1,2,3}^{4,5,6} &=& \delta\!\left(\B{k}_1+\B{k}_2+\B{k}_3 -\B{k}_4-\B{k}_5-\B{k}_6\right)\,, \BR{LIA1b}
   ^\L \widetilde{\D}_{\,1,2,3}^{\,4,5,6} &=&\frac {\kappa \L }{4\pi }\big[k_1 ^2+k_2 ^2+k_3 ^2-k_ 4^2-k_5 ^2-k_6 ^2\big ]\ .~~  
\end{eqnarray}
\ese
Explicit calculation of $^\Lambda \~{W}_{\,1,2,3}^{4,5,6}$   in Appendix~\ref{a:effW} shows that this is indeed the case: the contributions $^\L{W}_{\,1,2,3}^{4,5,6}$ and  $^\L Q_{\,1,2,3}^{4,5,6}$ in \Eq{tHamD}  cancel each other (see section 1 in Appendix~\ref{a:effW}).
(Such cancelation of complicated expressions was one of the tests of consistency and correctness of our calculations and
our
   \verb"Mathematica" code).
  The same is true for all the higher interaction coefficients: they must be zero within LIA.

Thus we need to study the first-order correction to the LIA, which for the interaction coefficient can be schematically represented as follows:
\BSE{tW1}\BEA{tW1a}
{^1 \~W}_{\,1,2,3}^{\,4,5,6}&=& \ {^1 W}_{\,1,2,3}^{\,4,5,6}\nonumber \\  
&&\!\! +\  {^1_1 Q}_{\,1,2,3}^{\,4,5,6}\   + \ {^1_2 Q}_{\,1,2,3}^{\,4,5,6} + \ {^1_3 Q}_{\,1,2,3}^{\,4,5,6}\,,~~~~~~
\BR{tW1b}
{^1_1 Q}&\sim & {\frac{{^\Lambda T} \otimes {^1 T}}{{^\Lambda\! \Omega}}} \,,\qquad
{^1_2 Q} \sim  \frac{{^1 T} \otimes {^\Lambda T}}{{^\Lambda\! \Omega}}\,, 
\BR{tW1c}
 {^1_3 Q}  &\sim& -\, {^1 \Omega}\, \frac{{^\Lambda T} \otimes {^\Lambda T}}{\left[\, {^\Lambda \Omega}\, \,\right]^2 }  \ .
\end{eqnarray}\ese
Here $ {^1 W}$ is the $\L^0$-order  contribution in the bare vertex $W$, given by \Eq{W-introC},
$ {^1 Q}$ is the $\L^0$-order  contribution in $Q$ which consists of
$^1_1 Q$ and $^1_2 Q$ originating from the part  $^1 T$ in the four-wave interaction coefficient $T$, and $^1_3 Q$
originating from the $^1\! \D$ corrections to the frequencies $\D$ in \Eqs{tHamE} and \eq{tHamF}.
Explicit \Eqs{Q}   for ${^1_1 Q}_{\,1,2,3}^{\,4,5,6}$, ${^1_2 Q}_{\,1,2,3}^{\,4,5,6}$  and  ${^1_3 Q}_{\,1,2,3}^{\,4,5,6}$  are presented in Appendix~\ref{aa:tW-LIA-1}. They are very lengthy and were analyzed using \texttt{Mathematica}, see \Sec{sss:an}.

\subsection{\label{ss:effD}Effective  six-KW dynamics}
\subsubsection{\label{sss:eq}Effective  equation of motion}
The LIA cancelation~\eq{LIA1a} on the full manifold~\eq{6-KW} is not exact:
\begin{eqnarray} 
  ^\Lambda \~{W}_{\,k,1,2}^{3,4,5}\  \d_{\,k,1,2}^{\,3,4,5}\ \d \big(\widetilde{\D} _{\,k,1,2}^{\,3,4,5} \big)\ne 0\,,~~~~~~~~~~~~\, \nonumber \\ 
  {\widetilde{\Omega}}_{\,k,1,2}^{\,3,4,5} \= \omega_k +\omega_1 +\omega_2 -\omega_3 -\omega_4 -\omega_5 \nonumber \\ 
  =  {^\Lambda\widetilde{\Omega}}_{\,k,1,2}^{\,3,4,5} +{^1\widetilde{\Omega}}_{\,k,1,2}^{\,3,4,5} +\mathcal{O}(\Lambda^{-1})\ . \nonumber
\end{eqnarray}
 The  residual  contribution due to  $^1\widetilde{\D} _{\,k,1,2}^{\,3,4,5}$ has to be accounted for -- an important fact overlooked in the previous KW literature, including the formulation of the effective KW dynamics recently presented by KS in
\cite{KS09}.
Now we are prepared to take another crucial step on the way to the effective KW model by replacing the
frequency $\o_k$ by its leading order (LIA) part~\eq{OkB} and simultaneously compensating for
the respective shift in the resonant manifold by correcting  the effective vertex $\~{\C H}$.
This corresponds to the following Hamiltonian equation,
\BEA{EdynA}i ~\frac{\p b_k}{\p t}&=&~^\L \o_k\, b_k \br
&&+ \frac{1}{12} \sum_{k+1+2=3+4+5}\!\!\!   {\C W}_{\,k,1,2}^{\, 3,4,5}\,b_1b_2b_3^* b_4^*b_5^*\,,
\eea
where the constraint $\B{k}+\B{k}_1+\B{k}_2=\B{k}_3+\B{k}_4+\B{k}_5$ holds.
 Here $\C W_{\,k,1,2}^{\,3,4,5}$ is a corrected interaction coefficient, which is calculated
  in  Appendix~\ref{App:cW}:
\BSE{Edyn} \BEA{effWa} 
    \C W_{\,k,1,2}^{\,3,4,5} &=& ^1 \~{ W}_{\,k,1,2}^{\,3,4,5} + ~^1\widetilde{S}_{\,k,1,2}^{\,3,4,5}\,, \BR{effWb}
^1 \widetilde{S}_{1,2,3}^{4,5,6} &=&
\frac{2\pi}{9  \kappa}\ {^1\widetilde{\Omega}_{1,2,3}^{4,5,6}}\!\!
\sum_{\substack{i=\{1,2,3\} \\ j=\{4,5,6\}}}\!\!\!\!\!
\frac{(\partial_{j} + \partial_{i})\, {^\Lambda\!\~W_{1,2,3}^{4,5,6}}}{(k_j-k_i)\ \Lambda}\,,~~~
\eea\ese
where $\partial_j (\cdot)\= \partial (\cdot)/\partial k_j$.

\eq{EdynA} represents a correct effective model and, will serve as a basis for our future analysis of KW dynamics and kinetics.
However, to make this equation useful  we need to complete the calculation of the effective interaction coefficient
 $  {\C W}_{\,k,1,2}^{\, 3,4,5}$ and simplify it to a reasonably tractable form. The key for achieving this is in a remarkably simple asymptotical behavior of $  {\C W}_{\,k,1,2}^{\, 3,4,5}$, which will be demonstrated in the next section.
 Such asymptotical expressions for $  {\C W}_{\,k,1,2}^{\, 3,4,5}$ will allow us to establish nonlocality of the KS theory,
 and thereby establish precisely {that} these asymptotical ranges with widely separated scales
 are the most dynamically active, which would lead us to the remarkably simple effective model expressed by
 the LNE ~\eq{LNE}.

\subsubsection{\label{sss:an} Analysis of the effective interaction coefficient $  {\C W}_{\,k,1,2}^{\, 3,4,5}$}

Now we will examine the asymptotical properties of the interaction coefficient which will be important for our study of locality of the KW spectra and formulation of the LNE ~\eq{LNE}.
The effective six-KW interaction coefficient ${\C W}_{\,k,1,2}^{\, 3,4,5}$ consists of five contributions given by ~\Eqs{tW1} and \eq{Edyn}. The explicit form of  ${\C W}_{\,k,1,2}^{\, 3,4,5}$ involves about $2 \times 10^4$ terms. However its asymptotic expansion in various regimes,    analyzed by \texttt{Mathematica}  demonstrates very clear and physical transparent behavior, which we will study upon the LIA resonance manifold
\BSE{LIAm}\BEA{LIAa}
\B k+\B k_1+\B k_2&=&\B k_3+\B k_4+\B k_5\,,
\BR{LIAb}k^2+k_1^2+k_2^2&=&k_3^2+k_4^2+k_5^2\ .
\eea \ese
  If the smallest wavevector
(say $\B k_5$)
is much smaller than the largest wave vector (say $\B k$)
 we have a remarkably simple expression:
\BEA{ASa}
  {\C W}_{\,k,1,2}^{\, 3,4,5} \to -\frac{3}{4 \pi\kappa} \B k \B k_1 \B  k_2 \B  k_3 \B  k_4 \B  k_5\,,
  \\
  \;\;\; \hbox{as} \;\;\; \frac{\min \{k, k_1, k_2, k_3, k_4, k_5\}}{\max \{k, k_1, k_2, k_3, k_4, k_5\}} \to 0\ .
  \nonumber
\eea
We emphasize that in the expression \eqref{ASa}, it is enough for the minimal wave vector to be much less
than the maximum wave number, and not all of the remaining five wave numbers in the sextet.
This was established using  \texttt{Mathematica} and Taylor expanding ${\C W}_{\,k,1,2}^{\, 3,4,5} $
with respect to one, two and four wave numbers~\footnote{The limit of three small wave numbers  is not allowed by the resonance conditions. Indeed, putting three wave numbers to zero, we get a $1 \leftrightarrow 2$ process which is not allowed in 1D for $^\Lambda \omega \sim k^2$.}. All of these expansions give the same leading term as in \eqref{ASa}, see Apps.~\ref{aa:W4} and \ref{aa:W2}.


     The form of  expression  \eqref{ASa}
   demonstrates a very simple physical fact: long KWs (with small $k$-vectors) can contribute to the energy of a  vortex line only when they produce curvature. The curvature, in turn,  is proportional to wave amplitude $b_k$ and, at a fixed amplitude, is inversely proportional to their wave-length, i.e. $\propto k$.  Therefore, in the effective motion equation each $b_j$ has to be accompanied by $k_j$, if $k_j\ll k$. This statement is exactly reflected by formula \eqref{ASa}.

Furthermore, a numerical evaluation of ${\C W}_{\,k,1,2}^{\, 3,4,5} $ on a set of $2^{10}$ randomly chosen wave numbers, different from each other at most by a factor of two, indicate that in the majority of cases its values are close to the asymptotical expression \eqref{ASa} (within 40\%).  Therefore, for most purposes we can approximate the effective six-KW interaction coefficient by the simple expression
\eqref{ASa}. Finally, our analysis of locality seen later in this paper, indicates that the most important wave sextets are
those which include modes with widely separating wavelengths, i.e. precisely those described by
the asymptotic formula \eqref{ASa}.

This leads us to the conclusion that the effective model for KW turbulence should use the
interaction coefficient \eqref{ASa}. Returning back to the physical space, we thereby obtain
the desired \em Local Nonlinear Equation \em (LNE) for KWs given by~\eqref{LNE}.

As we
mentioned in the beginning of the present paper, LNA is very close (isomorphous for small amplitudes) to
 the TLIA model \eqref{tLIA} introduced and simulated in \cite{boffetta}.
 It was argued in \cite{boffetta} that the TLIA model is a good alternative to the
original Biot-Savart formulation due to it dramatically greater simplicity. In the present paper we have found a further
support for this model, which is strengthened by the fact that now it follows from a detailed asymptotical
analysis, rather than being introduced \em ad hoc.\em

\subsubsection{\label{sss:sep}Partial contributions to the 6-wave effective interaction coefficient}
It would be instructive to demonstrate the relative importance of different partial contributions, $^1 W$, $^1_1 Q$, $^1_2 Q$, $^1_3 Q$ and ${^1 \widetilde{S}}$ [see \Eqs{tW1} and \eq{Edyn}] to the full  effective six-wave interaction coefficient. For this, we consider the simplest case, when four wave vectors are small,
say $k_1, k_2,k_3,k_5 \to 0$. We have (see Appendix \ref{aa:W2}):
 \BSE{sep}
\begin{eqnarray}
    \label{sepA}
\frac{^1 W}{\C{W}} &\to& \,-1 +\frac32\ln\!(k\ell)\, ~~~~~~~~~~~~~~ \,,    \\
    \label{sepB}
\frac{^1 _1 Q}{\C{W}} &\to& +\frac12 -\frac32\ln\!(k\ell) - \frac16\ln\!\frac{k_3}{k}\,,\\
    \label{sepC}
\frac{^1 _2 Q}{\C{W}} &\to& +\frac12 -\frac32\ln\!(k\ell) - \frac16\ln\!\frac{k_3}{k}\,,\\
    \label{sepD}
\frac{^1 _3 Q}{\C{W}} &\to& +1\, +\frac32\ln\!(k\ell) + \frac16\ln\!\frac{k_3}{k}\,,\\
    \label{sepE}
\frac{^1 \widetilde{S}}{\C{W}} &\to&  ~~~~~~~~~~~~~~~~~~~~~~\frac16\ln\!\frac{k_3}{k}\ .
\end{eqnarray}
\ese

One sees that \Eqs{sep} for the partial contributions  involve the artificial separation scale $\ell$, which cancels out from ${^1\~W} = {^1 W}+{^1 _1 Q}+{^1 _2 Q}+{^1 _3 Q}$.   This is not surprising because the initial expressions Eqs.~(\ref{W-intro}) do not contain $\ell$ but rather $\ln(1/a_0)$.
This cancelation serves as one more independent check of consistency of the entire procedure.

Notice  that  in the KS paper~\cite{KS04},   contributions~\eq{sepD} and \eq{sepE} were mistakenly not accounted for. Therefore the resulting KS expression for the six-wave effective interaction coefficient   depends on the artificial separation scale $\ell$.  This fact was missed  in
their numerical simulations~\cite{KS04}. In their recent paper~\cite{KS09}, the lack of contribution~\eq{sepD} in the previous work was acknowledged
(also in \cite{boffetta}), but the contribution \eq{sepE}
was still missing.


\section{\label{s:kin} Kinetic description of KW turbulence}

\subsection{\label{ss:KE}Effective Kinetic Equation   for KWs}
The statistical description of weakly interacting waves can be reached~\cite{ZLF} in terms of the kinetic equation (KE) shown below for the continuous limit $k\mathcal{L}\gg 1$,
\BSE{KE}\BE{KEa}
\p n(\B k, t)\big / \p t= \mbox{St}( \B  k, t)\,,
\ee
for the spectra $n(\B  k, t)$ which are  the simultaneous pair correlation functions, defined by
\BE{KEb}
    \< b(\B  k, t)b^*(\B  k',t)\> = \frac{2\pi}  {\mathcal{L}}\, \delta(\B k-\B k') \, n( \B  k, t)\,,
\ee
where $\< \dots \>$ stands for proper (ensemble, etc.) averaging. In the classical limit \footnote{Here, we evoke a quantum mechanical analogy
as an elegant shortcut, allowing us to introduce KE and the respective solutions easily. However, the reader should not get confused with this
analogy and  understand that our KW system is purely classical. In particular, the Plank's constant $\hbar$ is irrelevant outside of this
analogy, and should be simply replaced by 1.},
when the occupation numbers of Bose particles $N(\B  k, t)\gg 1$, $n(\B  k, t)=\hbar N(\B   k, t)$,
the collision integral St$(\B   k, t)$ can be found in various ways~\cite{KS04,LNR-1,ZLF}, including the Golden Rule of quantum mechanics. For the  $3\leftrightarrow 3$ process of KW scattering, described by the motion \Eq{EdynA}:
\bea\nn
\mbox{St}_{3\leftrightarrow 3}(\B   k)  & =& \frac{\pi}{12}\!\int\!\!\!\!\int\!\!\!\!\int\!\!\!\!\int\!\!\!\!\int \Big|\C W_{\,\,  k, 1,2}^{\,3,4,5}\Big|^2  \, \d_{\,k, 1,2}^{\,3,4,5} \  \d \Big( {^\L \D_{\,k, 1,2}^{\, 3,4,5}} \Big)\br
&& \hskip -1 cm \times \big(n_k^{-1}+ n_1^{-1}+n_{2}^{-1}-n_{3}^{-1}-n_{4}^{-1}-n_{5}^{-1} \big)\BR{KEc}
  && \hskip -1 cm \times \,  n_{ k}n_1n_2n_3n_4n_5 \  d \B  k_1\, d \B  k_2\, d \B  k_3\, d \B  k_4\,  d  \B k_5 \ .
\eea\ese
KE~\eq{KE} conserves the total number of (quasi)-particles $N$ and the total (bare) energy of the system $~^\L E$, defined respectively as follows:
\BE{NE} N\= \int n_k \, d\B k\,, \quad  ~^\L E\= \int ~^\L \o_k n_k \, d\B  k \ .
\ee
KE~\eq{KE} has a Rayleigh-Jeans solution,
\BE{sol} n\Sb {T}(k)=\frac {T}{\hbar ~^\L \o_k + \mu}\,,
\ee
which corresponds to thermodynamic equilibrium of KWs with temperature $T$ and chemical potential $\mu$.

In various wave systems, including the KWs described by KE~\eq{KEa},  there also exist flux-equilibrium solutions, $n\Sb E (k)$ and $n\Sb N (k)$,  with constant  $k$-space fluxes of energy and particles respectively. The corresponding solution  for  $n\Sb E (k)$  was suggested in the KS-paper~\cite{KS04} under an (unverified) assumption of  locality of the $E$-flux.  In \Sec{ss:nonloc}, we will analyze this assumption in the framework of the derived KE~\eq{KE}, and will prove that it is wrong.
The $N$-flux solution $n\Sb N (k)$ was discussed in \cite{inverseKW}. In  \Sec{ss:nonloc}, we will show that this spectrum is marginally nonlocal, which means that it can be ``fixed" by a logarithmic correction.

\subsection{\label{ss:nonloc} Phenomenology of the E- and N-flux equilibrium solutions for KW turbulence}
Conservation laws~\eq{NE} for $E$ and $N$ allow one to introduce the continuity equations for $n_k$ and $^\L E_k\= ~^\L \o_k n_k$ and their corresponding fluxes in the $k$-space, $\mu_k$ and $\ve_k$:
\BSE{cont}
\BEA{contA}
  \frac{\p\, n_k}{\p t}+ \frac{\p \mu_k}{\p\B  k }=0\,, && \mu_k\= - \int_0^k \mbox{St}_{3\leftrightarrow 3}( \B  k)\,d \B k\,,  
\BR{contB}
  \frac{\p\, {^\L\! E_k}}{\p t}+ \frac{\p \ve_k}{\p \B k }=0\,,   &&
   \ve_k\= - \!\int_0^k \!\!\! {~^\L \o_k\, \mbox{St}_{3\leftrightarrow 3}(\B   k)\,d \B k}\ .\ \ \ ~~~~~
\eea\ese
In  scale-invariant systems, when the frequency and interaction coefficients are homogeneous functions of wave vectors, \Eqs{cont}  allow one
   to guess the scale-invariant  flux equilibrium solutions of KE~\eq{KE}~\cite{ZLF}:
\BE{si}n \Sb E (k) = A\Sb E k^{-x \Sb E}\,, \quad n\Sb N (k) = A\Sb N k^{-x\Sb N}\,,
\ee
Here $A\Sb E $ and
$A\Sb N$  are some dimensional constants. Scaling exponents  $x\Sb N$ and $x\Sb E$ can  be found in the case of \emph{locality} of the $N$- and $E$-fluxes, i.e. when the integrals over   $\B k_1,\dots \B  k_5$ in \Eqs{cont} and~\eq{KEc} converge. In this case, the leading contribution to these integrals originate from regions where $k_1\sim k_2\sim k_3\sim k_4\sim k_5\sim k$ and thus, the fluxes~\eq{cont} can be estimated as follows:
\BSE{flux}
\BEA{fluxA}
\mu_k& \simeq& k^5 [\C W(k,k,k|k,k,k)]^2 n\Sb{N}^5 (k)\Big / \o_k \,,
\BR{fluxB}
\ve _k& \simeq& k^5 [\C W(k,k,k|k,k,k)]^2 n\Sb{N}^5 (k)  .
\eea\ese
Stationarity of  solutions of \Eqs{cont} require constancy of their respective fluxes: i.e. $\mu_k$ and $\ve_k$ should be $k$-independent. Together with \Eqs{flux} this allows one to find the scaling exponents in \Eq{si}.

Our formulation~\eq{KE} of KW kinetics belongs to the scale-invariant  class \footnote{It is evident for the approximation \Eq{ASa}. For the full expression \Eq{effWa} it was confirmed by symbolic computation with the help of \texttt{Mathematica}.}: 
\begin{eqnarray}
    && ^\L \o_k\propto k^{\,2}\,,\ \mbox{and for}\ \forall\, \eta   \nonumber \\
    && {\C W}(\eta \B k, \eta \B k_1,\eta \B k_2\,|\,\eta \B k_3,\eta \B k_4,\eta \B k_5) \nonumber \\ 
    = {\eta}^6\!\!\!\!\!\! && {\C W}(\B  k, \B  k_1, \B k_2\,|\, \B k_3, \B k_4, \B k_5)\ . \nonumber 
\end{eqnarray}
Estimating $\C W(k,k,k|k,k,k)\simeq k^6/\kappa$ and $^\L \o_k\simeq \kappa\L k^2$ in \Eqs{flux}, one gets for $N$-flux  spectrum~\cite{inverseKW}:
\BSE{sol2}
\BE{sol2A}  n\Sb {N} (k) \simeq     \big( \mu \kappa\big / \L \big) ^{1/5} \, k^{-3}\,, \quad x\Sb N = 3 \,,
\ee
and for $E$-flux KS-spectrum~\cite{KS04}:
\BE{sol2B}
n\Sb {E} (k) \simeq      \big ( \ve\kappa^2 \big) ^{1/5}\,  k^{-17/5} \,, \quad  x\Sb E = 17/5\ .
\ee\ese

\subsection{\label{ss:nonloc}  Non-locality of the $N$- and $E$-fluxes  by $3   \leftrightarrow 3$-scattering}
Consider  the $3 \!\leftrightarrow \!3$ collision term~\eq{KEc} for KWs with the interaction
amplitude  $  \C W_{\,1,2,3}^{\,4,5,6}$. Note that in~\eq{KEc}   $\int d \B k_j$ are
one-dimensional integrals $\int_{-\infty}^ \infty d \B k_j $.
Let us examine  the ``infrared" (IR) region ($\,k_5\ll k, k_1, k_2, k_3, k_4\,$)
in the integral ~\eq{KEc}, taking into account the asymptotics~\eq{ASa},
and observing that the expression
\bea
  && \d_{\,k, 1,2}^{\, \,3,4,5} \d \big( ^\L \~\D_{\,k, 1,2}^{\, \,3,4,5} \big) \big(n_k^{-1}\!+ n_1^{-1}\!+n_{2}^{-1}\!-n_{3}^{-1}\!-n_{4}^{-1}\!-n_{5}^{-1} \big)
  \nn \cr
&&  ~~~~~~~~~~~~~~~~~~~ \times\,  n_{ k}\,n_1\,n_2\,n_3\,n_4\,n_5\br \nn \cr
&&   \to
    \d_{\,k, 1,2}^{\, \,3,4}
   \d \big( ^\L\! \D_{\,k, 1,2}^{\, \,3,4} \big) \big(n_k^{-1}+ n_1^{-1}+n_{2}^{-1}-n_{3}^{-1}-n_{4}^{-1} \big)
   \nn \cr
&& ~~~~~~~~~~~~~~~~~~~~~~ \times\,  n_{ k}\,n_1\,n_2\,n_3\,n_4\,n_5 \sim n_5 \sim k_5^{\,-x}.
  \eea
  Thus  the integral over $\B k_5$ in the IR region can be factorized and written as follows:
  \BE{int}
2 \int _0
   k_5^2\,  n(k_5\!)\,  d  k_5 \propto 2  \int _{1/\ell}  k_5^{2-x} d k_5 \ .
\ee
The factor 2 here originates from the symmetry of the integration area and evenness of the integrand: $\int_{-\infty}^\infty= 2 \int _0 ^\infty$.   The lower limit $0$ in
  this expression should be replaced by the smallest wave number where
   the assumed scaling behavior~\eq{si} holds, and moreover, it depends on the particular way the wave system is forced.
   For example, this cutoff wave number could be $1/\ell$,
   where $\ell$ is  the  mean inter-vortex separation $\ell$, at which one expects a cutoff of the wave spectrum.
   The crucial assumption of locality, under which both the $E$-flux (KS) and the $N$-flux spectra were obtained, implies that
   the integral ~\eq{int} is independent of this cutoff in the limit $\ell \to 0$.
Clearly, integral~\eq{int} depends on the IR-cutoff if $x\ge 3$, which is the case for  both the  $E$-flux (KS) and the $N$-flux spectra~\eq{sol2}.    Note that all other integrals over $\B k_1$, $\B k_2$, $\B k_3$ and $\B k_4$ in \eq{KEc} diverge exactly in the same manner as the integral
over $\B k_5$, i.e. each of them leads to expression ~\eq{int}.

Even stronger IR divergence occurs when two wave numbers on the same side of the sextet (e.g. $\B k_1$ and $\B k_2$, or $\B k_3$ and $\B k_4$, etc.)
are small. In this case, integrations over both of the small wave numbers will lead to the same contribution, namely integral~\eq{int},
i.e. the result will be  the integral~\eq{int} squared.
This appears to be the strongest IR singularity, and the  resulting behavior of the collision integral
\Eq{KEc} is
 \BE{int2}
\mbox{St}\Sp{IR} \sim
  \left( \int _{1/\ell}  k_5^{2-x} d k_5 \right)^2 .
\ee
 When two wave numbers from the opposite sides of the sextet (e.g. $\mathbf{k}_2$ and $\mathbf{k}_5$) tend to zero simultaneously, we get an extra small factor in the integrand because in this case  $\big(n_k^{-1}+ n_1^{-1}-n_{3}^{-1}-n_{4}^{-1} \big) \to 0$. As a result we get IR convergence
  in this  range. One can also show IR convergence when two wave numbers from one side and one on the other side of the sextet are small
 (the resulting integral is IR convergent for $x<9/2$).

  Divergence of integrals in \Eq{KEc} means that both spectra~\eq{sol2} with $x\Sb N=3$ and  $x\Sb E=17/5>3$, obtained under \emph{opposite} assumption of the convergence of these integrals  in the limit $\ell\to \infty$ are not solutions of the  $3 \!\leftrightarrow \!3$-KE~\eq{KEc} and thus cannot be realized in Nature. One should find another self-consistent solution of this KE. Note, that the proof of divergence at the IR limits is sufficient for discarding the spectra under the test, whereas proving  convergence would require considering all the singular limits including the ultra-violet (UV) ranges.
  However, we have examined these limits, too. At the UV end we have  obtained convergence for the KS and for the inverse cascade spectra. Thus the most dangerous singularity appears to be in the IR range, when two wave numbers from the same side of the wave sextet are small simultaneously.

\subsection{\label{sss:log} Logarithmic corrections for the $N$-flux spectrum~\eq{sol2A}}
Note that for the $N$-flux spectrum~\eq{sol2A} $n\Sb N (k)\propto k^{-3}$, that the integrals~\eq{int} and  \eq{int2} diverge only logarithmically.  The same situation happens, e.g. for the direct enstrophy cascade in two-dimensional turbulence: dimensional reasoning leads to the Kraichnan-1967~\cite{Kra67} turbulent energy spectrum
\BSE{Kra}\BE{KraA}
E(k)\propto k^{-3}
\ee
for which the integral for the enstrophy flux diverges logarithmically. Using a simple argument of constancy of the enstrophy flux, Kraichnan suggested~\cite{Kra71} a logarithmic correction to the spectra
\BE{KraB}
E(k)\propto k^{-3} \ln^{-1/3}(k l)\,, 
\ee \ese
that permits the enstrophy flux to be $k$-independent. Here $l$ is the enstrophy pumping scale.

Using the same arguments, we can substitute in \Eq{KEc}, a logarithmically corrected spectrum $n\Sb N (k)\propto k^{-3} \ln^{-x}(k \ell)$ and find $x$ by the requirement that the resulting $N$-flux, $\mu_k$, \Eq{contA} will be $k$-independent. Having in mind that according to \Eq{fluxA} $\mu_k\propto n\Sb N^5$, we can guess that  $x=1/5$. Then, the divergent integral~\eq{int} will be $\propto \ln ^{4/5}(k\ell)$, while the remaining convergent integrals in \Eq{KEc} will be $\propto \ln ^{-4/5}(k\ell)$. Therefore, the  resulting flux $\mu_{k}$ will be $k$-independent as it should be~\cite{Kra71}.  So, our prediction is that instead of a non-local spectrum~\eq{sol2A} we have a slightly steeper log-corrected spectrum
\BE{corr}
n\Sb N (k) \simeq     \frac{\big (\mu\, \kappa \big) ^{1/5}}{k^{3}\ln^{1/5} (k\, \ell )}\ .\ee
The difference is not large, but the underlying physics must be correct;  as one says on the Odessa market: ``We can argue the price, but the weight must be correct".

\section*{Conclusions}

In this paper, we have derived an effective theory of KW turbulence based on asymptotic expansions of the Biot-Savart model in
powers of small $1/\L$ and small nonlinearity,
by applying a canonical transformation eliminating non-resonant  low-order (quadric) interactions, and by using
the standard Wave Turbulence approach based on  random phases~\cite{ZLF}.
In doing so, we have fixed errors arising from the previous derivations, particularly the latest one by KS ~\cite{KS09}, by taking into
account previously omitted and important contributions to the effective six-wave interaction coefficient.
We have examined the resulting six-wave interaction coefficient in several  asymptotic limits when one or several
wave numbers are in the IR range. These limits are summarized in a remarkably simple expression  \eqref{ASa}.
This allowed us to achieve three goals:
\begin{itemize}
\item
To derive a simple effective model for KW turbulence expressed in the Local Nonlinear Equation \eqref{LNE}.
In addition to small $1/\L$ and the weak nonlinearity, this model relies on the fact that our findings show, for dynamically relevant wave sextets, the interaction coefficient is a simple product of the
six wave numbers,  \Eq{ASa}.
For weak nonlinearities, the LNE is isomorphic to the previously suggested  TLIA model \cite{boffetta}.
\item
To examine the locality of the $E$-flux (KS) and the $N$-flux spectra. We found that the KS spectrum
is non-local and therefore cannot be realized in Nature.
\item The $N$-flux spectrum is found to be
marginally non-local and could be ``rescued" by a logarithmic correction, which we constructed following a
qualitative Kraichnan approach. However, it remains to be seen if such a spectrum can be realized
in Quantum Turbulence, because, as it was shown in \cite{KW_rec}, the vortex line reconnections
can generate only the forward cascade and not the inverse one (i.e. the reconnections produce an effectively
large-scale wave forcing).
\end{itemize}

Finally we will discuss the numerical studies of KW turbulence.
The earliest numerics  by KS were reported in ~\cite{KS04}. They claimed that they observed the KS spectrum.
At the same time they gave a value of the $E$-flux constant $\sim 10^{-5}$ which is unusually small.
We have already mentioned that this work failed to take into account several important contributions to the
effective interaction coefficient, and thus these numerical results cannot be trusted.
In particular, we showed that their interaction coefficient must have contained a spurious dependence
on the scale $\ell$ which makes the numerical results arbitrary and dependent on the choice of such a cutoff.
In addition, even if the interaction coefficient was correct, the Monte-Carlo method used by KS is
a rather dangerous tool when one deals with slowly divergent integrals (in this case $\int_0 x^{-7/5} \, dx$).

On the other hand,  recent numerical simulations of the TLIA model also reported agreement with
the KS scaling (as well as an agreement with the inverse cascade scaling) \cite{boffetta}.
How can one explain this now when we showed analytically that the KS spectrum is non-local?
It turns out that the correct KW spectrum, which takes into account the non-local interactions with
long KW's, has an index which is close (but not equal) to the KS index, and it is also consistent
with the data of \cite{boffetta}. We will report these results in a separate publication.

\acknowledgements
We are very grateful to Mark Vilensky for fruitful discussions and help.
We acknowledge the support of the U.S. - Israel Binational
Science Foundation,   the support of the European Community --
Research Infrastructures under the FP7 Capacities Specific Programme,
MICROKELVIN project number 228464.

\appendix

\section{\label{a:BIC} Bare interactions}
\subsection{\label{aa:calc} Actual calculation of the bare interaction coefficients}
The geometrical constraint of a small amplitude
perturbation can be expressed in terms of a parameter
\begin{equation}
\epsilon(z_1,z_2)=|w(z_1)-w(z_2)|/|z_1 - z_2| \ll 1.
\end{equation}
This allows one to expand Hamiltonian~\eq{BSE} in powers of $\epsilon$ and to re-write it in terms consisting of the number of wave interactions, according to \Eq{exp}. KS found the exact expressions for $H_2$, $H_4$ and $H_6$ \cite{KS04}:
\begin{eqnarray}\label{eq:KS04Ham}
{H}_2&=&\frac{\kappa}{8\pi}\int \frac{d z_1 d z_2}{|z_1-z_2|}\left[2Re\left(w^{'*}(z_1)w^{'}(z_2)\right)-\epsilon^2\right],\br
{H}_4&=&\frac{\kappa}{32\pi}\int \frac{d z_1 d z_2}{|z_1-z_2|}\left[3\epsilon^4-4\epsilon^2Re\left(w^{'*}(z_1)w^{'}(z_2)\right)\right],\\
{H}_6&=&\frac{\kappa}{64\pi}\int \frac{d z_1 d z_2}{|z_1-z_2|}\left[6\epsilon^4Re\left(w^{'*}(z_1)w^{'}(z_2)\right)-5\epsilon^6\right]\nonumber.
\end{eqnarray}

The explicit calculation of these integrals was analytically done in \cite{boffetta}, by evaluating the terms in (\ref{eq:KS04Ham}) in Fourier space, and then expressing each integral as various cosine expressions \cite{KS04}.  Hamiltonian (\ref{eq:KS04Ham}) can be expressed in terms of a wave representation variable $a_k=a(\B k,t)$ by applying a Fourier transform~\eq{Four} in the variables $z_1$ and $z_2$, (for details see \cite{KS04,boffetta}). The result is given by \Eqs{Ham}, in which the cosine expressions for $\omega_k$, $T_{12}^{34}$ and $W_{123}^{456}$ were done in \cite{KS04}. In our notations  they  are
\begin{widetext}
\bea
\omega_k  &=&  \frac{\kappa}{2\pi}\left[A-B\right]\,,   \qquad
T_{12}^{34}=\frac{1}{4\pi}\left[6D-E\right] \,, \qquad
W_{123}^{456}=\frac{9}{4\pi\kappa}\left[3P-5Q\right]\,, \qquad \mbox{where}~~~~~\label{KST}
 \br
A &=& \int_{a_0}^\infty \frac{dz_-}{z_-}k^2C^k\,, \quad
B = \int_{a_0}^\infty \frac{dz_-}{z^3_-}\big[1-C^k\big]\,,  \quad
D =\int_{a_0}^\infty \frac{dz_-}{z_-^5}\big[1-C_1-C_2-C^3-C^4+C^3_2+C^{43}+C^4_2\big]\,, \br
E &=& \int_{a_0}^\infty \frac{dz_-}{z_-^3}\big[k_1k_4\left(C^4+C_1-C^{43}-C^4_2\right)+k_1k_3\left(C^3+C_1-C^{43}-C^3_2
\right) +k_3k_2\left(C^3+C_2-C^{43}-C^3_1\right)\label{eq:D}\\ 
&&~~~~~~~~~~~~ +k_4k_2\left(C^4+C_2-C^{43}-C^3_2\right)\big], \br
P &=& \int_{a_0}^\infty\frac{dz_-}{z_-^5}k_6k_2[C_2-C^5_2-C_{23}+C^5_{23}-C^4_2+C^{45}_2+C^4_{23}
-C^6_1+C^6-C^{56}- C^6_3+C^{56}_3-C^{46}+C^{456}+C^{46}_3-C_{12}]\,, \br
Q &=& \int_{a_0}^\infty\frac{dz_-}{z_-^7}\big[1-C^4-C_1+C^4_1-C^6+C^{46}+C^6_1-C^{46}_1-C^5 +C^{45}
+C^5_1-C^{45}_1+ C^{65}
-C^{456}  -C^{56}_1+C_{23}\br 
&&~~~~~~~~~~~~ -C_3+C^4_3+C_{13}
-C^4_{13}+C^6_3-C^{46}_3-C^6_{13}+C^5_2
+ C^5_3-C^{45}_3-C^5_{13}+C^6_2-C^{56}_3+C_{12}+C^4_2-C_2\big]\ . \label{eq:Q}
\end{eqnarray}
\end{widetext}
Here the variable, $z_- = |z_1-z_2|$ and the expressions $C$, are cosine functions such that $C_1 = \cos(\B k_1z_-)$, $C^4_1 = \cos((\B k_4-\B k_1)z_-)$, $C^{45}_1=\cos((\B k_4+\B k_5-\B k_1)z_-)$, $C^{45}_{12}=\cos((\B k_4+\B k_5-\B k_1-\B k_2)z_-)$ and so on.  The lower limit of integration ${a_0}$ is the induced cutoff of the vortex core radius ${a_0}<|z_1-z_2|$.

The trick  used for explicit calculation of  the analytical form of these integrals was suggested and used in ~\cite{boffetta}. First  one should integrate by parts all the cosine integrals, so they can be expressed in the form of $\displaystyle
\int^\infty_{a_0} \frac{\cos(z)}{z}dz $.
Then, one can use a cosine identity for this integral \cite{GR80},
\begin{eqnarray}\label{eq:cosineformula}
&& \int_{a_0}^\infty\frac{\cos(z)}{z}dz = -\gamma-\ln({a_0})-\int_0^{a_0}\frac{\cos(z)-1}{z}dz~~~~\br &&= -\gamma-\ln({a_0})-\sum_{k=1}^\infty\frac{\left(-{a_0^2}\right)^k}{2k\left(2k\right)!}= -\gamma-\ln(|{a_0}|)+\mathcal{O}({a_0^2})\,,
\end{eqnarray}
where $\gamma=0.5772\dots$ is  the Euler Constant. Therefore, in the limit of a small vortex core radius ${a_0}$, we can neglect terms of order $\sim {a_0^2}$ and higher.  For example, let's consider the following general cosine expression that can be found in \Eqs{eq:D}: $\displaystyle
\int^\infty_{a_0}\!\! z^{-3}\,{\cos(\mathcal{K}z)}dz$,
where $\mathcal{K}$ is an expression that involves a linear combination of wave numbers, i.e. $\mathcal{K}= k_1-k_4$.  Therefore, integration by parts will yield the following result for this integral:
\begin{eqnarray*}
&& \int^\infty_{a_0} \frac{\cos(\mathcal{K}z)}{z^3}dz= \left[-\frac{\cos(\mathcal{K}z)}{2z^2}\right]^\infty_{a_0}
+\left[\frac{\mathcal{K}\sin(\mathcal{K}z)}{2z}\right]^\infty_{a_0}\\
&&-\frac{\mathcal{K}}{2}\int^\infty_{a_0} \frac{\cos(\mathcal{K}z)}{z}dz\\
&=&\frac{\cos(\mathcal{K}{a_0})}{2{a_0^2}}-\frac{\mathcal{K}\sin(\mathcal{K}{a_0})}{2{a_0}}
-\frac{\mathcal{K}^2}{2}\int^\infty_{\mathcal{K}{a_0}}\frac{\cos(y)}{y}dy\ .
\end{eqnarray*}
We then expand $\cos(\mathcal{K}{a_0})$ and $\sin(\mathcal{K}{a_0})$ in powers of ${a_0}$, and apply the cosine formula (\ref{eq:cosineformula}) for the last integral, where in the last step we have also changed integration variables: $y=\mathcal{K}z$.  The final expression is then
\begin{eqnarray*}
\int\limits ^\infty_{a_0} \frac{\cos(\mathcal{K}z)}{z^3}dz = \frac{1}{2{a_0^2}}-\frac{3\mathcal{K}^2}{4} + \frac{\mathcal{K}^2}{2}\left[\gamma + \ln(|\mathcal{K}{a_0}|)\right]+\mathcal{O}({a_0^2})\ .
\end{eqnarray*}
By applying a similar procedure to the other cosine integrals, we find that all terms of negative powers of ${a_0}$, (that will diverge in the limit ${a_0}\to 0$) actually cancel in the final expression for each interaction coefficient.   Applying this strategy to all interaction cofficients,  we get the following   analytical evaluation of the Hamiltonian functions~\cite{boffetta}:
\begin{eqnarray}
&&\qquad\quad \Lambda_0 = \ln(\ell/a_0)\,, \nonumber \\  
\omega_k &=& \frac{\kappa k^2}{4\pi}\Big[\L_0 -\gamma -\frac{3}{2}- \ln ( k\ell )\Big] ,\label{res1}\br
T_{12}^{34} &=& \frac{1}{16\pi}\left[\B k_1\B k_2\B k_3\B k_4 (1+4\gamma-4\L_0 ) -\mathcal{F}_{\,1,2}^{\,3,4}\right]\,,  \br
W_{123}^{456}&=& \frac{9 }{32\pi\kappa}\left[\B k_1\B k_2\B k_3\B k_4\B k_5\B k_6 (1-4\gamma+4\L_0 ) -\mathcal{G}_{\,1,2,3}^{\,4,5,6}\right]\ .
\end{eqnarray}
Explicit equations for $\C F_{12}^{34}$ and $\C G_{123}^{456}$ are given below in Appendices \ref{aa:F4} and \ref{aa:G6}. In the main text we introduced $\Lambda \= \Lambda_0 -\gamma - 3/2$. Writing $\Lambda = \ln(\ell/a)$, we see that $a = a_0 e^{\gamma+3/2} \simeq 8 a_0$.

\subsection{\label{aa:F4} Bare 4-wave interaction function  $\mathcal{F}_{\,1,2}^{\,3,4}$}
A rather cumbersome calculation, presented above, results  in an explicit equation for the  4-wave interaction function  $\mathcal{F}_{\,1,2}^{\,3,4}$ in \Eqs{res1} and \eq{T-introB}.
Function $\mathcal{F}_{\,1,2}^{\,3,4}$ is a symmetrical version of $F_{\,\,1,2}^{\,3,4}$:  $\mathcal{F}_{\,1,2}^{\,3,4}\= \left\{F_{\,\,1,2}^{\,3,4}\right\}\sb{S}$ where the  operator $\left\{\dots\right\}\sb{S}$ stands for the symmetrization $\B k_1 \leftrightarrow \B k_2$, $\B k_3 \leftrightarrow \B k_4$ and $\{\B k_1,\B k_2\} \leftrightarrow \{\B k_3,\B k_4\}$. In its turn $F_{\,\,1,2}^{\,3,4}$ is defined as following:
\bse
\begin{eqnarray}
    \label{def-of-F}
  F_{\,\,1,2}^{\,3,4} &\equiv&  \sum_{\C K \in \C K_1}{\C K^4}\ln\!\left({|\C K|}{\ell}\right) \label{WFd}\br &&  +2\sum_{i,j}\sum_{\C K \in \C K_{ij}}k_ik_j\,{\C K^2}\ln\!\left({|\C K|}{\ell}\right)\ .
\end{eqnarray}
The $\sum_{i,j}$ denotes sum of four terms with  $(i,j)=\big\{(4,1), (3,1),(3,2),(4,2)\big\} $, $\C K$ is either a single wave vector or linear combination of wave-vectors that belong to one of the following   sets:
\bea\nn
   \C K_1&=&\left\{^ -[_1],^-[_2],^-[^3],^-[^4],^+[^3_2],^+[^{43}],^+[^4_2] \right\},\\ \label{WF1a}
   \C K_{41}&=& \left\{^+[^4], ^+[_1], ^-[^{43}],^-[^4_2]\right\}\,,\br
   \C K_{31} &=& \left\{^+[^3], ^+[_1], ^-[^{43}],^-[^3_2]\right\} \,,  \\ \nn
   \C K_{32}&=&\left\{^+[^3], ^+[_2], ^-[^{43}],^-[^3_1]\right\}\,, \br
   \C K_{42}&=&\left\{^+[^4], ^+[_2], ^-[^{43}],^-[^4_1]\right\} \ .
\eea
\ese
Here we used the following shorthand notations with $\a\,, \b\,, \g=1,2,3,4$:
$\displaystyle
[^\alpha]\= \B k_\a\,, \ [_\beta]\= -\B k_\beta\,, \  [^\a_\b]\=\B  k_\a -\B k_\b\,, \ [^{\a\g}]\= \B k_\a+\B k_\g\,, \ 
[_{\b\g}]\= -\B k_\b-\B k_\g\,,
$
and $~^+ $ or $~^- $ signs before $[\dots]$ should be understood as prefactors $+1$ or $-1$ in the corresponding term in the sum. For example:
\begin{eqnarray*}
    \nonumber
    && {\C K^4}\ln\!\left({|\C K|\ell}\right) \ \mbox{for}\ \C K\in \{^-[ _1]\} \ \mbox{is}\ -{k_1^4}\ln\!\left({k_1\ell}\right), \\ 
    \nonumber
    && {\C K^4}\ln\!\left({|\C K|\ell}\right) \ \mbox{for}\ \C K\in \{^+[ ^4_2]\} \ \mbox{is}\ +{(\B k_4-\B k_2)^4}\ln\! \left({|\B k_4-\B k_2|\ell}\right),\\
    \nonumber
    &&\B  k_i\B k_j\,{\C K^2}\ln\!\left({|\C K|\ell}\right) \ \mbox{for}\ i=4,\, j=1,\, \\ &&  \C K\in \{^-[ {^{43}}]\} \ \mbox{is} ~-\B k_4\B k_1\,{(\B k_4+\B k_3)^4} \ln\! \left({|\B k_4+\B k_3|\ell}\right)\, .
\end{eqnarray*}

\subsection{\label{aa:G6} Bare 6-wave interaction function  $\mathcal{G}_{\,\,1,2,3}^{\,4,5,6}$}
Function $\mathcal{G}_{\,1,2,3}^{\,4,5,6} \= \left\{G_{\,1,2,3}^{\,4,5,6}\right\}\sb{S}$. The operator $\left\{\dots\right\}\sb{S}$ stands for the symmetrization $\B k_1 \leftrightarrow \B k_2 \leftrightarrow \B k_3$, $\B k_4 \leftrightarrow \B k_5 \leftrightarrow \B k_6$ and $\{\B k_1,\B k_2,\B k_3\} \leftrightarrow \{\B k_4,\B k_5,\B k_6\}$, and $G_{\,1,2,3}^{\,4,5,6}$ is defined as following:

\bse \label{def-of-G} \begin{eqnarray}
   G_{\,1,2,3}^{\,4,5,6} &\equiv& \sum_{\C K \in \C K_3} \B k_6\B k_2\,{\C K^4} \ln\!\left({|\C K|\ell}\right) \nonumber \\ && +\frac{1}{18}\sum_{\C K \in \C K_4}{\C K^{\,6}} \ln\!\left({|\C K|\ell}\right)\,, 
\end{eqnarray}
where
\begin{eqnarray}
   \C K_3 &=& \Big\{ {^+}[_2],{^-}[^5_2],{^-[_{23}]},{^+}[^5_{23}],{^-}[^4_2],{^+}[^{45}_2],{^+}[^4_{23}],{^-}[^6_1],{{^+}}[^6], \nonumber \\ 
   && {^-}[^{56}],{^-}[^6_3],{^+}[^{56}_3],{^-}[^{46}],{^+}[^{456}],{^+}[^{46}_3],{^-}[_{12}]\Big\},\\ 
   \C K_4 &=& \Big\{ {{^-}}[{^4}],{^-}[_1],{^+}[^4_1],{^-}[^6],{^+}[^{46}],{^+}[^6_1],{^-}[^{46}_1],{^-}[^5],\nonumber \\ 
   && {^+}[^{45}],{^+}[^5_1], {^-}[^{45}_1],{^+}[^{65}],{^-}[^{456}],{^-}[^{56}_1],{^+}[_{23}],{^-}[_3],\nonumber \\ 
   && {^+}[^4_3],{^+}[_{13}],{^-}[^4_{13}],{^+}[^6_3],{^-}[^{46}_3],{^-}[^6_{13}],{^+}[^5_2],{^+}[^5_3],\nonumber \\ 
   && {^-}[^{45}_3],{^-}[^5_{13}],{^+}[^6_2],{^-}[^{65}_3],{^+}[_{12}],{^+}[^4_2],{^-}[_2]\Big\}.
\end{eqnarray}

\ese

\section{\label{a:effW}Effective six-KW interaction coefficient}

\subsection{\label{aa:Q}Absence of 6-wave dynamics in LIA}
According to \Eqs{tHamD} and \eq{tHamE}, the expression for $^\Lambda\~ W_{\,\,1,2,3}^{\,4,5,6}$ is given by
\BSE{tHam-app}
\BEA{tHamA-app}
  {^\Lambda{\~W}_{\,1,2,3}^{4,5,6}}   &=&  {^\Lambda W_{\,1,2,3}^{4,5,6}} + {^\Lambda Q_{\,1,2,3}^{4,5,6}}\,, \BR{tHamE-app}
   {^\Lambda Q_{\,1,2,3}^{4,5,6}}  &=& \frac{1}{8}\!\! \sum^{3}_{\substack{i,j,m=\,1\\ i\neq j\neq m}} \sum^{6}_{\substack{p,q,r=\,4\\ p\neq q \neq r}}\!\!\! {^\Lambda q_{\,\,i,j,m}^{\,p,q,r}}\,, \br  
   {^\Lambda q_{\,\,i,j,m}^{\,p,q,r}} &\=&  \frac{^\Lambda T_{\,r,\,j+m-r}^{j,\,m} \, {^\Lambda T_{\,i,\,p+q-i}^{q,\,p}}}{ ^\Lambda \Omega_{\,j,\,m}^{\,r,\,j+m-r}} \nonumber \\ 
   && +\frac{^\Lambda T_{\,m,\,q+r-m}^{q,\,r} \, ^\Lambda T_{\,p,\,i+j-p}^{i,\,j}}{^\Lambda  \Omega_{\,q,\,r}^{\,m,\,q+r-m}} \,,
\eea
\ese
where $ {^\Lambda \Omega}_{1,2}^{3,4} \= {^\Lambda\omega_1}+{^\Lambda\omega_2} -{^\Lambda\omega_3}-{^\Lambda\omega_4} $. We want to compute this equation on the LIA manifold~\eq{LIAm}. To do this we express two  wave vectors in terms of the other four \footnote{It is appropriate to remind the reader, that we use the bold face notation of the one-dimensional wave vector for convenience only. Indeed, such a vector is just a real number, $\B k \in \mathbb{R}$.} using the LIA manifold constraint~\eq{LIAm}:
\bse
\label{k1k4-LIA}
\begin{eqnarray}
    \B k_1 &=& \frac{\left(\B k_3-\B k\right) \left(\B k_2-\B k_3\right)}{\B k+\B k_2-\B k_3-\B k_5}+\B k_5\,, \\ 
    \B k_4 &=& \frac{\left(\B k_3-\B k\right) \left(\B k_2-\B k_3\right)}{\B k+\B k_2-\B k_3-\B k_5}+\B k+\B k_2-\B k_3\ .
\end{eqnarray}
\ese
Then ${^\Lambda{\~W}_{\,1,2,3}^{4,5,6}}$ is easily simplified to zero with the help of \texttt{Mathematica}. This gives an independent verification of the validity of our initial \Eqs{tHam} for full interaction cofficient $^\Lambda\~W_{\,\,1,2,3}^{\,4,5,6}$ which is needed for the calculations of the $\C O(1)$ contribution ${^1 \~W_{\,\,1,2,3}^{\,4,5,6}}$. Another way to see the cancelation is to use  the Zakharov-Schulman variables~\cite{ZS82} that parameterise the LIA manifold~\eq{LIAm}.

\subsection{\label{aa:tW-LIA-1}Exact expression for ${^1 \~W}$}

We get expressions for $^1_1 Q$, $^1_2 Q$ and $^1_3 Q$, introduced by \Eqs{tW1}, from \Eqs{tHamD} and \eq{tHamE}. Namely:
\begin{widetext}
\BSE{Q}
  \begin{equation} \label{Qa}
^1_1Q_{1,2,3}^{4,5,6} = \frac{1}{8}\!\!\sum^{3}_{\substack{i,j,m=\,1\\ i\neq j\neq m}} \sum^{6}_{\substack{p,q,r=\,4\\ p\neq q \neq r}} \Bigg[{{\frac{^\Lambda T_{\,r,\,j+m-r}^{j,\,m} \, {^1 T_{\,i,\,p+q-i}^{q,\,p}}}{ ^\Lambda \Omega_{\,j,\,m}^{\,r,\,j+m-r}}  +\frac{^\Lambda T_{\,m,\,q+r-m}^{q,\,r} \, ^1 T_{\,p,\,i+j-p}^{i,\,j}}{^\Lambda \Omega_{\,q,\,r}^{\,m,\,q+r-m}}}}\Bigg ] \,, \ 
\end{equation}
 \begin{equation}
\label{Qb}
^1_2 Q_{1,2,3}^{4,5,6} = \frac{1}{8}\!\!\sum^{3}_{\substack{i,j,m=\,1\\ i\neq j\neq m}} \sum^{6}_{\substack{p,q,r=\,4\\ p\neq q \neq r}} \Bigg[{{\frac{^1 T_{\,r,\,j+k-r}^{j,\,k} \, {^\Lambda T_{\,i,\,p+q-i}^{q,\,p}}}{ ^\Lambda \Omega_{\,j,\,m}^{\,r,\,j+m-r}}  +\frac{^1 T_{\,m,\,q+r-m}^{q,\,r} \, ^\Lambda T_{\,p,\,i+j-p}^{i,\,j}}{^\Lambda \Omega_{\,q,\,r}^{\,m,\,q+r-m}}}}\Bigg ] \,,  
\end{equation}
 \begin{equation}
\label{Qc}
   ^1_3 Q_{1,2,3}^{4,5,6} =  
  \frac{1}{8} \sum^{3}_{\substack{i,j,m=\,1\\ i\neq j\neq m}} \sum^{6}_{\substack{p,q,r=\,4\\ p\neq q \neq r}}  \Bigg [ \frac{^\Lambda T_{\,r,\,j+m-r}^{j,\,m} \, {^\Lambda T_{\,i,\,p+q-i}^{q,\,p}}}{\big(\, ^\Lambda \Omega_{\,j,\,m}^{\,r,\,j+m-r}\,\big)^2} \cdot {^1 \Omega_{\,j,\,m}^{\,r,\,j+m-r}}  +\,\frac{^\Lambda T_{\,m,\,q+r-m}^{q,\,r} \, ^\Lambda T_{\,p,\,i+j-p}^{i,\,j}}{\big(\,^\Lambda \Omega_{\,q,\,r}^{\,m,\,q+r-m}\,\big)^2} \cdot {^1 \Omega_{\,q,\,r}^{\,m,\,q+r-m}} \Bigg ] .~~~~~
\end{equation}
\ese
\end{widetext}
Again, using \texttt{Mathematica} we substitute \Eqs{k1k4-LIA} into \Eqs{Qa} -- \eqref{Qc}. Clearly, the resulting equations are too cumbersome to be presented here. But we will analyze them in various limiting cases, see below.

\subsection{\label{App:cW}Derivation of \Eq{effWb} for ${^1\! \widetilde{S}_{\,k,1,2}^{\,3,4,5}}$}

First of all, let us find a parametrization for the full resonant manifold, by calculation
of the correction to the LIA parametrization \eqref{k1k4-LIA},
namely
\bea
\B k_1 &=& {^\Lambda\!\B k_1} + {^1\!\B k_1}\,,
\qquad
\B k_4 = {^\Lambda\!\B k_4} + {^1\!\B k_4}\ ,
\eea
where $^\Lambda\!\B k_1$ and $^\Lambda\!\B k_4$
are given by the right-hand sides of \Eqs{k1k4-LIA} respectively.
Corrections  $^1\!\B k_1$ and $^1\!\B k_4$ are found so that the
resonances in $\B k$, \Eq{LIAa}, and (full) $\omega$ are satisfied.
The resonances in $\B k$ fixes ${^1\!\B k_1} = {^1\!\B k_4}$.
Then the $\omega$-resonance in the leading order in
$1/\Lambda$ gives
\bea
\widetilde{\Omega}_{1,2,3}^{4,5,6} &=& {^1\!\B k_1}
\frac{\partial {^\Lambda \omega_1}}{\partial \B{k}_1}
- {^1\!\B k_4} \frac{\partial {^\Lambda\! \omega_4}}{\partial \B k_4} \nonumber \\
&&\!\!+\ {^1\widetilde{\Omega}_{1,2,3}^{4,5,6}} +\mathcal{O}(\Lambda^{-1}) =0\ .
\eea
Thus
\be
{^1\!\B k_1} = {^1\!\B k_4} \approx
\frac{2\pi}{\Lambda \kappa} \, \frac{^1 \widetilde{\Omega}_{1,2,3}^{4,5,6}}{(\B k_4-\B k_1)}\ .
\ee
This allows us to write down the contribution of $^\Lambda\!\~W$ from the deviation of the LIA resonant surface:
\bea
^1 \widetilde{S}_{1,2,3}^{4,5,6} &=&
{^1\!\B k_1} \, \frac{ \partial \,  ^\Lambda\!\~W_{1,2,3}^{4,5,6}}
{\partial \B k_1}
+ {^1\!\B k_4} \frac{ \, \partial \,  ^\Lambda\!\~W_{1,2,3}^{4,5,6}}
{\partial \B k_1} +\mathcal{O}(\Lambda^{-1}) \nonumber \\
&\approx&
\frac{2\pi}{\Lambda \kappa} \, ^1\!\widetilde{\Omega}_{1,2,3}^{4,5,6} \,
\frac{(\partial_{4} + \partial_{1}) \,  ^\Lambda\!\~W_{1,2,3}^{4,5,6} }{(\B k_4-\B k_1)}\,,
\eea
with $\partial_{j}(\cdot) = \partial_{j}(\cdot)/\partial \B k_j$.
It is obvious that instead of $\B k_1$ and $\B k_4$ we could use  parametrizations
in terms of other pairs  $\B k_i$ and $\B k_j$ with
$i=1,2$ or 3 and $j=4,5$ or 6. This enables us to write
a fully symmetric expression for $^1 S$:
\be
^1 \widetilde{S}_{1,2,3}^{4,5,6} =
\frac{2\pi}{9 \Lambda \kappa} \, {^1\!\widetilde{\Omega}_{1,2,3}^{4,5,6}} \!
\sum_{\substack{i=\{1,2,3\} \\ j=\{4,5,6\}}}\!\!\!\!
\frac{(\partial_{j} + \partial_{i}) \,  ^\Lambda\!\~W_{1,2,3}^{4,5,6} }{(\B k_j-\B k_i)}.
\ee
This is the required expression \Eq{effWb}.

\subsection{\label{aa:W4} Analytical expression for $\C W$ on the LIA manifold when
two wave numbers are small}

 Let us put together the coefficients to the interaction coefficient ${\C W}_{\,k,1,2}^{\, 3,4,5}$ given
 in~\eq{tW1a}, \eq{tW1b}, \eq{tW1c}, \eq{W-introC} and \eq{Edyn}, and use in these
 expressions the formulae obtained in the previous appendices
 and the parametrization of the LIA surface (\ref{k1k4-LIA}). Using
   \texttt{Mathematica}, and Taylor expanding ${\C W}_{\,k,1,2}^{\, 3,4,5} $
with respect to  one wave number, e.g.
$\B k_5$, we obtain a remarkably simple result, - expression  \eqref{ASa}.

Now we will consider the asymptotical limit when two of the wave numbers, say $\B k_2$ and
$\B k_5$ (let them be on the opposite sides of the resonance conditions), are much less
than the other wave numbers in the sextet.
 Using
   \texttt{Mathematica} and Taylor expanding ${\C W}_{\,k,1,2}^{\, 3,4,5} $
with respect to  two wave numbers $\B k_2$ and
$\B k_5$, we have
\BEA{W4-in-lim}   
     \lim_{\scriptsize \begin{array}{c}
          \B  k_2 \to 0  \\
          \B  k_5 \to 0
         \end{array}}\!\!
     {\C W}_{\,k,1,2}^{\, 3,4,5} &=&  -\frac{3}{4\pi\kappa}k^2\B k_2 k_3^2\B k_5\,,
\eea
Simultaneously, see Eq.~(\ref{k1k4-LIA}):
\begin{eqnarray}
    \label{k1k4-Limits}
      \lim_{\scriptsize \begin{array}{c}
          \B  k_2 \to 0  \\
          \B  k_5 \to 0
         \end{array}}\B  k_1  &\to& \B k_3\,, \quad 
      \lim_{\scriptsize \begin{array}{c}
          \B  k_2 \to 0  \\
          \B  k_5 \to 0
         \end{array}}\B  k_4  \to \B k\ .
\end{eqnarray}
Therefore, \eqref{W4-in-lim} coincides with
\eqref{ASa}. Note that this was not obvious \em a priori\em, because formally \eqref{ASa} was obtained
when $\B k_5$ is much less than the rest of the wave numbers, including $\B k_2$.

For reference, we provide expressions for the different contributions  to the interaction coefficient ${\C W}_{\,k,1,2}^{\, 3,4,5}$ given  in Eqs.~\eq{tW1} and  (\ref{Edyn}). For $\B k_2, \B k_5 \to 0$:
\begin{widetext}
\bse
\begin{eqnarray}
{^1 W} &\to& -\frac{3}{4 \pi\kappa}k^2 \B k_2  k_3^2 \B k_5
        \bigg[+\frac32\ln\!{(k\ell)} -\frac{1}{24}\Big(49-\frac{(1-x)^2 (7+10x+7 x^2)}{x^2}\ln|1-x|\nonumber \\
        &&~~~~~~~~~~~~~~~~~~~~~~ +2 x (12+7 x) \ln|x|-7\frac{(1+x)^4}{x^2}\ln|1+x|\Big)\bigg]\,,\\
{^1 _1 Q} = {^1 _2 Q} &\to& -\frac{3}{4 \pi\kappa}k^2 \B k_2  k_3^2 \B k_5
        \bigg[-\frac32\ln\!{(k\ell )} +\frac{1}{48}\Big(59 -\frac{(1-x)^2 (9+10x+9 x^2)}{x^2}\ln|1-x|\nonumber \\
        &&~~~~~~~~~~~~~~~~~~~~~~ +2\big(9 x^2+14x-6+\frac{2}{1-x}\big) \ln |x| -9\frac{(1+x)^4}{x^2}\ln|1+x|\Big)\bigg]\,,\\
{^1 _3 Q} &\to& -\frac{3}{4 \pi\kappa}k^2 \B k_2  k_3^2 \B k_5
        \bigg[+\frac32\ln\!{(k\ell )} +\frac{1}{48}\Big(7+\frac{(1-x)^2 \left(1+x^2\right)}{x^2}\ln|1-x|\nonumber \\
        &&~~~~~~~~~~~~~~~~~~~~~~ +2\frac{1-5 x+x^3}{1-x}\ln |x| +\frac{(1+x)^4}{x^2}\ln|1+x|\Big)\bigg]\,,\\
{^1 S} &\to& -\frac{3}{4 \pi\kappa}k^2 \B k_2 k_3^2 \B k_5
        \left[\,\frac{1}{6}\,\frac{1+x}{1-x}\,\ln |x|\right] \,,\\
\nonumber \\
{^1 \~W} &\to& -\frac{3}{4 \pi\kappa}k^2 \B k_2 k_3^2 \B k_5
        \left[\,1 -\frac16\,\frac{1+x}{1-x}\,\ln|x|\,\right]\,, \qquad \qquad x \= \B k_3/\B k \ .
\end{eqnarray}
\ese
\end{widetext}
Another possibility is for two small wave numbers to be on the same side of the sextet.  We have checked that on the resonant manifold, this also leads to \eqref{ASa}.

\subsection{\label{aa:W2} Analytical expression for $\C W$ on the LIA manifold when
four wave numbers are small}

Now let us, using \texttt{Mathematica},
calculate the asymptotic behavior of $\C W$ when four wave vectors
are smaller than the other two; on the LIA manifold this automatically simplifies to $k_1,k_2, k_3, k_5 \ll k,k_4$ (remember
that on the LIA manifold $\B k_1$ and $\B k_4$ are expressed in terms of the other wave numbers using \Eq{k1k4-LIA}, thus from (\ref{k1k4-Limits})) we have
\BEA{W2-in-lim}   
    \lim_{\scriptsize
           k_{\,1,2,3,5}\, \to\, 0}
     {\C W}_{\,k,1,2}^{\, 3,4,5}
   &=& -\frac{3}{4 \pi\kappa}\,k^2 \B k_2 k_3^2 \B k_5.
\eea
Again, we have got an expression which  coincides with
\eqref{ASa}. We emphasize  that this was not obvious \em a priori\em, because formally \eqref{ASa} was obtained
when $k_5$ is much less than the rest of the wave numbers, including $k_1, k_2, k_3$.

{\textit{Therefore we conclude that the expression \eqref{ASa} is valid when $k_5$ is much less than just
one other wave number in the sextet, say $k$, and not only when it is much less than all of the
remaining wave numbers.}}

For a reference, we give the term by term results for the limit
 $k_1,k_2, k_3, k_5 \ll k,k_4$:
 \begin{eqnarray*}
{^1 W} &\to& -\frac{3}{4 \pi\kappa}k^2 \B k_2 k_3^2 \B k_5
        \left[\, -1 +\frac32\ln\!{(k\ell)}\,+~~0~~~~\right], \nonumber\\
{^1 _1 Q} &\to& -\frac{3}{4 \pi\kappa}k^2 \B k_2 k_3^2 \B k_5
        \left[\,+\frac12 -\frac32\ln\!{(k\ell)} -\frac16\ln\!\frac{k_3}{k}\,\right],\nonumber\\
{^1 _2 Q} &\to& -\frac{3}{4 \pi\kappa}k^2 \B k_2 k_3^2 \B k_5
        \left[\,+\frac12 -\frac32\ln\!{(k\ell)} -\frac16\ln\!\frac{k_3}{k}\,\right],\nonumber\\
{^1 _3 Q} &\to& -\frac{3}{4 \pi\kappa}k^2 \B k_2 k_3^2 \B k_5
        \left[\,+1 +\frac32\ln\!{(k\ell)} +\frac16\ln\!\frac{k_3}{k}\,\right],\nonumber\\
{^1 S} &\to& -\frac{3}{4 \pi\kappa}k^2 \B k_2 k_3^2 \B k_5
        \left[\,0 ~~~~~+ ~~ 0  ~~~~ \, +\frac16\ln\!\frac{k_3}{k}\,\right]. \end{eqnarray*}
        The sum of this contributions is very simple:
$$\
{^1 \~W}  \to  -\frac{3}{4 \pi\kappa}k^2 \B k_2 k_3^2 \B k_5
        \left[\,+1 ~~~+ ~~0~~~~~ + ~~~0 ~ \right]  .
        $$


\end{document}